\documentclass[12pt, preprint]{aastex}
\newcommand\ea{et al.\ }
\newcommand\hst{{\it HST}}
\def\chandra{{\it Chandra}}
\newcommand\psc{\ifmmode{\rm\,cm^{-2}}\else{${\rm\,cm^{-2}}$}\fi}
\shortauthors{Levenson et al.}

\begin{document}
\title{Deconstructing NGC 7130}

\author{N. A. Levenson\altaffilmark{1}, K. A. Weaver\altaffilmark{2,3}, T. M. Heckman\altaffilmark{3}, H. Awaki\altaffilmark{4}, and Y. Terashima\altaffilmark{5}}
\altaffiltext{1}{Department of Physics and Astronomy, University of Kentucky,
Lexington, KY 40506; levenson@pa.uky.edu}
\altaffiltext{2}{Code 662, NASA/GSFC, Greenbelt, MD 20771}
\altaffiltext{3}{Department of Physics and Astronomy, Bloomberg Center, Johns Hopkins University, Baltimore, MD 21218}
\altaffiltext{4}{Department of Physics, Faculty of Science, Ehime University,
Bunkyo-cho, Matsuyama, Ehime 790-8577, Japan}
\altaffiltext{5}{Institute of Space and Astronautical Science, 3-1-1 Yoshinodai, 
Sagamihara, Kanagawa 229-8510, Japan}

\begin{abstract}
Observations of the Seyfert 2 and starburst galaxy NGC 7130 with the
{\it Chandra X-ray Observatory} illustrate that both of these phenomena
contribute significantly to the galaxy's detectable X-ray emission.
The active galactic nucleus (AGN) is strongly obscured, buried beneath
column density $N_H > 10^{24}\psc$, and it is most evident in a
prominent Fe K$\alpha$ emission line with equivalent width greater than
1 keV.  The AGN accounts for most (60\%) of the observed 
X-rays at energy E $>$ 2
keV, with the remainder due to spatially extended star formation.  The
soft X-ray emission is strong and predominantly thermal, on both small and large
scales.  We attribute the thermal emission to stellar processes.
In total, the AGN is responsible
for only one-third of the observed 0.5--10 keV luminosity of 
$3 \times 10^{41} {\rm \, erg\, s^{-1}}$ of
this galaxy, and less than half of its bolometric luminosity. 
Starburst/AGN composite galaxies like NGC 7130 are truly common, and similar
examples may contribute significantly to the high-energy
cosmic X-ray background
while remaining hidden at lower energies, especially if they
are distant.
\end{abstract}

\section{Introduction}

Accretion onto supermassive black holes powers active galactic nuclei
(AGNs), but star formation also may be fundamentally
related to the central black holes and their activity.
For example, galaxies with AGN tend to contain a significant
population of recently-formed stars, and the AGN luminosity
is correlated with the stellar age \citep{Kau03}.
The presence of 
large reservoirs of gas and the dynamical conditions for
star formation simultaneously fulfill the requirements for fueling an
active nucleus.  The underlying connection between
AGN and starbursts may be more significant than merely coincidental, being
truly causal.
In this interpretation, 
the black holes that are a consequence of massive star evolution 
serve as the seeds of the central supermassive black hole,
with subsequent stellar outflows and supernovae promoting accretion
to increase mass \citep[e.g.,][]{Wee83,Nor88}. 

Locally, transient bursts of star formation are correlated with
AGN.  Among the well-studied Seyfert 2s, the lower-luminosity
obscured variety of AGN, nearly half also
contain circumnuclear starbursts \citep*{Cid98,Gon01}.
These are the starburst/AGN composite galaxies.
While either star formation or accretion may dominate the
luminosity and classification of a particular galaxy,
the commonplace composite galaxies
demonstrate that the two processes are not exclusive.
Rather, both phenomena are often energetically
relevant and should be considered
together in  observations of real galaxies.

One  complication of 
the known starburst/AGN composite galaxies is that they
are preferentially more obscured than their counterparts
that lack starbursts \citep*{LWH01j,Ris99}.
In the extreme case where the obscuring column density 
$N_H > 1.5\times 10^{24} \psc$,
which are classified as Compton thick, 
the AGN X-ray emission is detected only indirectly
\citep*[e.g.,][]{Kro94,Ghi94}.
Although the
X-rays of a less-obscured AGN provide an unambiguous signature of 
the central engine's intrinsic power,
the signal of a Compton thick AGN is diminished by orders of magnitude, 
and the corresponding
starburst can be strong, particularly at soft X-ray energies ($E < 2$ keV).

In nearby galaxies, however, the confusing multiple
emission sources can be disentangled.
With high signal-to-noise and high spatial resolution observations,
the physical conditions 
can be determined directly and accurately.
Such detailed measurements are impossible in
more distant cases or surveys.
The 3\arcsec{} spectroscopic 
fibers of the Sloan Digital Sky Survey \citep{Yor00}, 
for example,
correspond to a  physical scale larger than 1 kpc at distances greater
than 70 Mpc.
Nearby case studies therefore serve as the empirical building
blocks of more distant and less well-studied galaxies.

We investigate the case of
NGC 7130, an example of a starburst/AGN composite galaxy.
Classified on the basis of optical
emission line ratios, it is a normal Seyfert 2 \citep*{Phi83}, while
most of the ultraviolet emission is  spectrally 
characteristic of star formation \citep{Thu84}.
The  circumnuclear starburst  is powerful and compact (90 pc),
evident in both the 
vacuum ultraviolet spectrum, which shows absorption features formed
in the winds and photospheres of massive stars, and in the optical spectrum,
where the high-order Balmer series and He I lines are observed in absorption
\citep{Gon98}.
Reprocessed starlight is evident in the far-infrared,
where NGC 7130 has been observed with the {\it Infrared Astronomical Satellite}
({\it IRAS}) and 
the {\it Infrared Space Observatory} \citep*{Spi02}.
NGC 7130 is also a known X-ray source \citep{Ris99},
with  spatial extent
and thermal emission indicating that 
the AGN alone is not responsible for the high-energy luminosity
\citep*{LWH01s}.
We illustrate here that both the starburst and the AGN
are energetically important, even in the X-ray regime.  

In this work, we present new observations of NGC 7130 obtained
with the {\it Chandra X-ray Observatory} (\chandra; \S\ref{sec:obs}).
Both the starburst and AGN characteristics are evident in
X-ray imaging and spectroscopy (\S\ref{sec:spec}).  
The broader spectral energy distribution of NGC 7130, particularly
at far-infrared wavelengths, reveals the relative bolometric
contributions of star formation and the AGN separately (\S\ref{sec:sed}).
The example of NGC 7130 demonstrates the difficulty of diagnosing
AGN properties when a starburst is present, particularly when 
sensitivity is limited (\S\ref{sec:impl}).  
For $H_0=70 {\rm \, km\,s^{-1}\,Mpc^{-1}}$, the distance to NGC 7130
is 69 Mpc, and $1\arcsec \equiv 330{\rm \, pc}$.

\section{Observations and Image Analysis\label{sec:obs}}
The \chandra{} Advanced CCD Imaging Spectrometer (ACIS)
observed NGC 7130 on the back-illuminated S3 detector
on 2001 October 23--24.
(See \citealt{Wei00} 
for more information on \chandra{}.) 
We reprocessed all data from original Level 1 event files 
removing the spectral and  spatial randomization that is included in standard
processing with \chandra{} Interactive Analysis of Observations (CIAO)
software. 
We applied the current gain corrections for the S3 CCD (from 
2001 July 31), and 
we included only good events that 
do not lie on node boundaries, where discrimination of cosmic rays
is difficult.
We examined the lightcurves of background regions and found no
significant flares, so we used all data within the standard 
good-time intervals, for a total exposure of 38.6 ks. 

\begin{figure}
\includegraphics[width=3.2in]{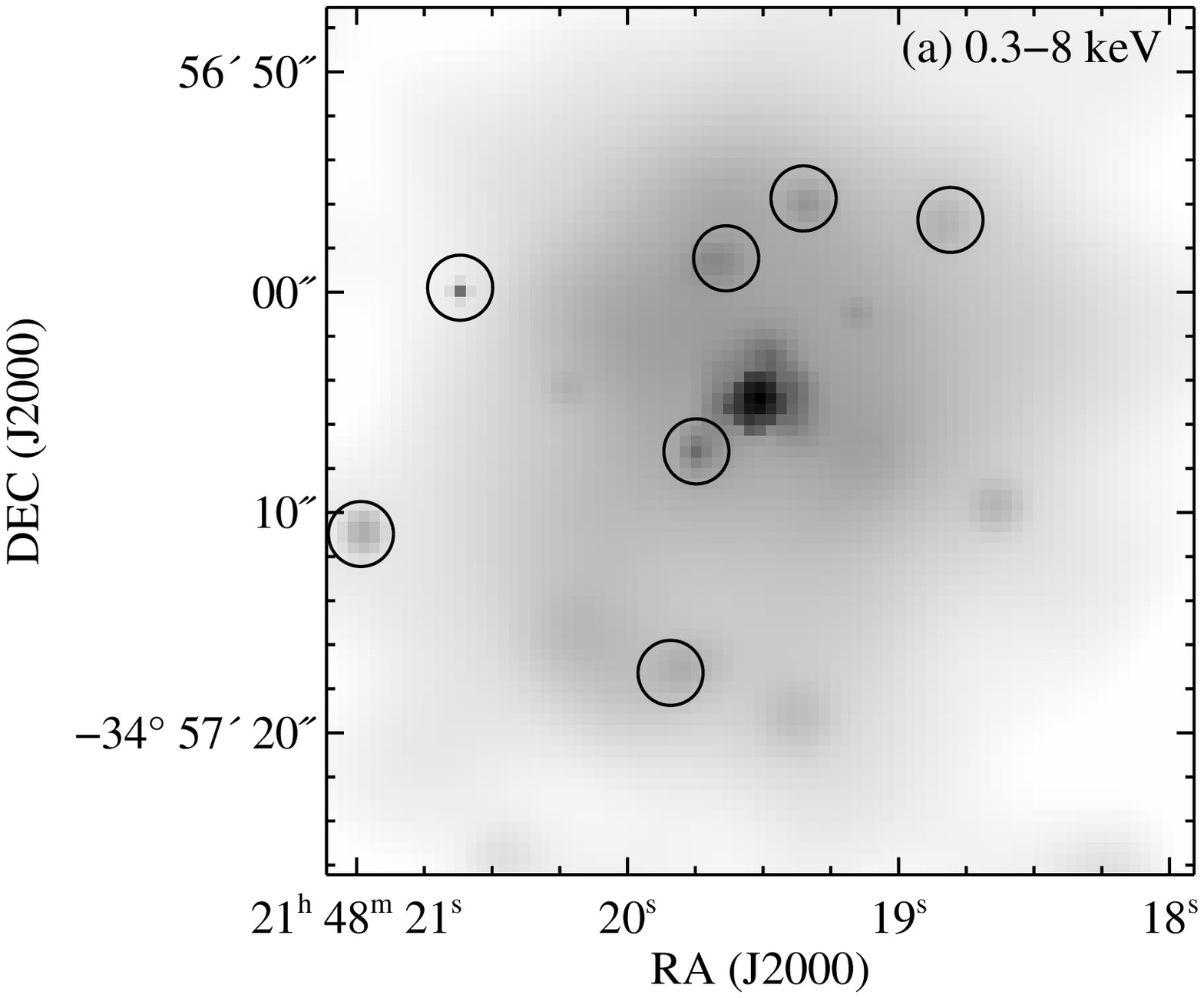} 
\includegraphics[width=3.2in]{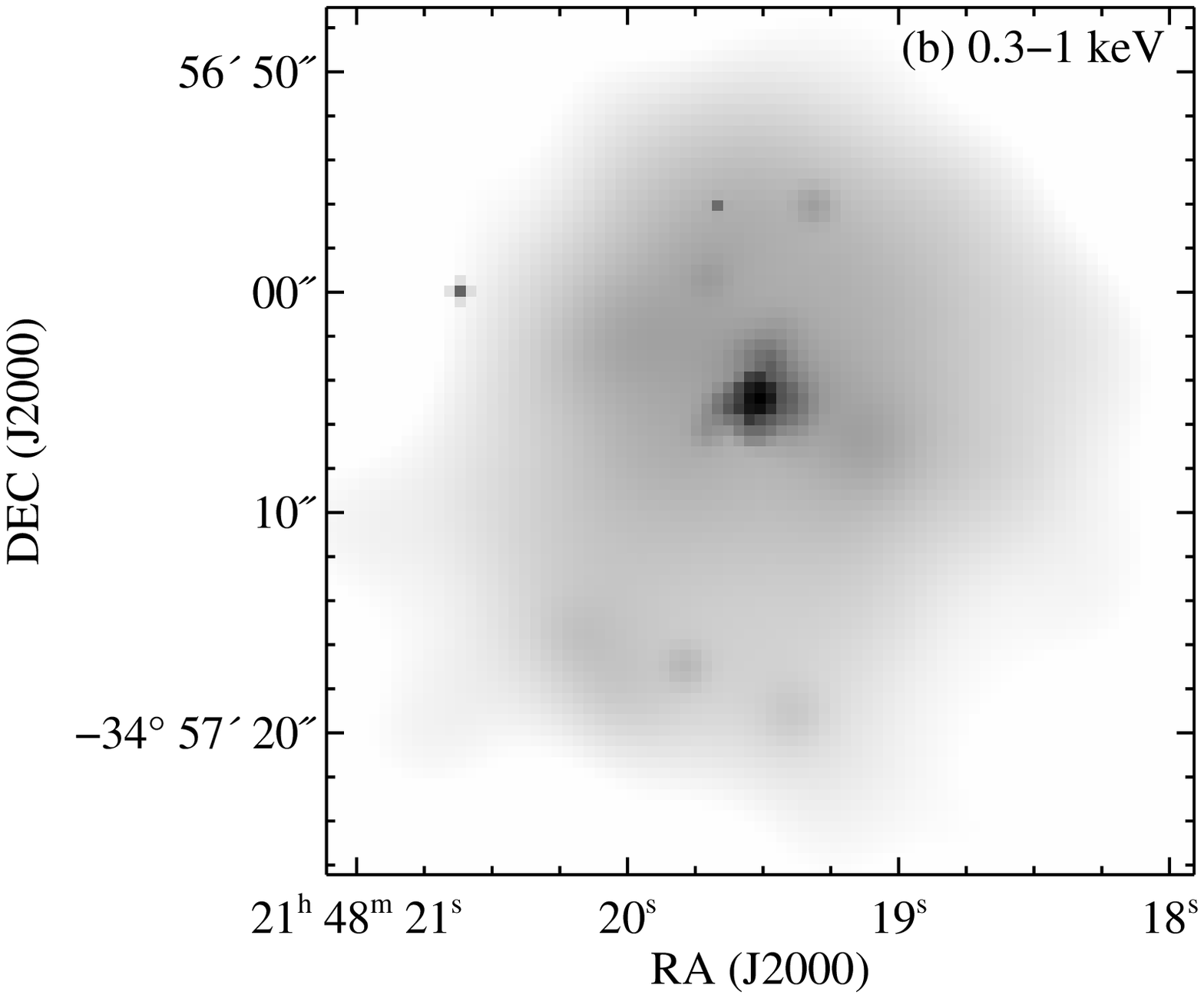} 

\includegraphics[width=3.2in]{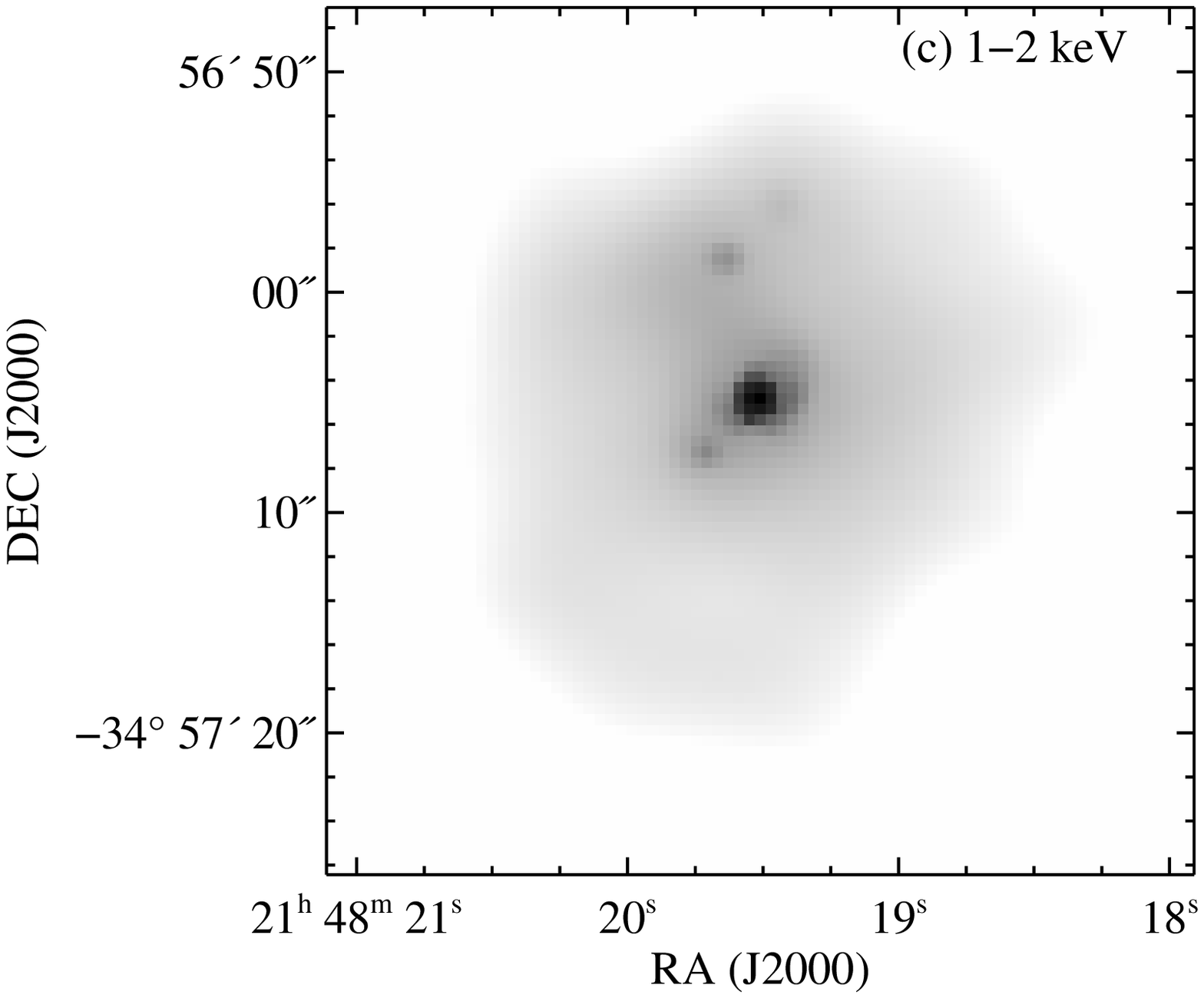} 
\includegraphics[width=3.2in]{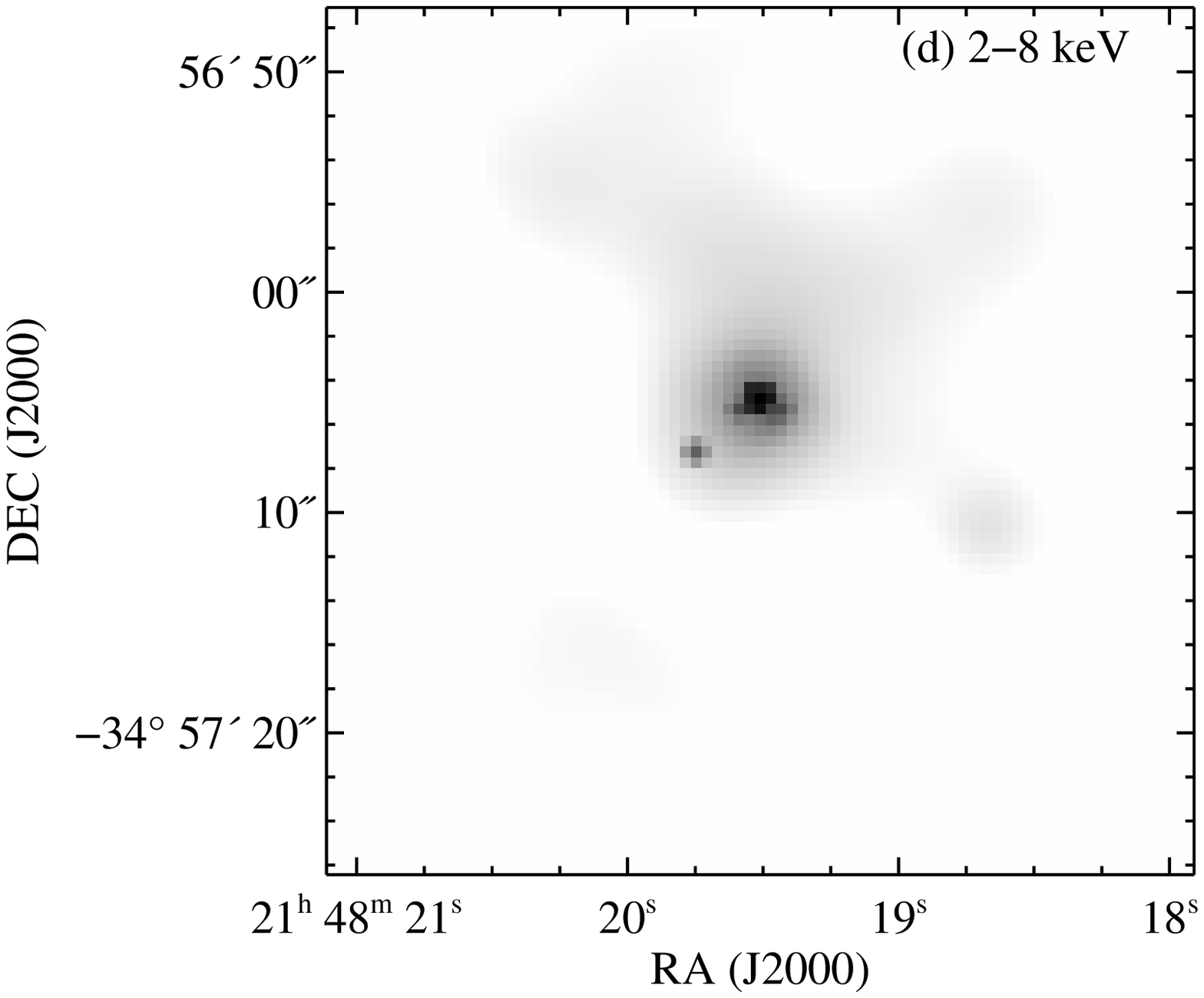} 
\caption{\label{fig:xrimg}({\it a}) Broad-band (0.3--8 keV) X-ray image of
NGC 7130.  The strong central concentration is due to both the AGN
and the circumnuclear starburst, while the very extended emission
is due to stellar processes alone. The additional
extranuclear sources are marked.
This image has been adaptively smoothed 
and is scaled logarithmically. 
({\it b}) \chandra{} soft X-ray (0.3--1 keV) image of NGC 7130,
 adaptively smoothed and scaled
logarithmically.  
({\it c}) \chandra{} medium X-ray (1--2 keV) image of NGC 7130,
 adaptively smoothed and
scaled logarithmically.  
({\it d}) \chandra{} hard X-ray (2--8 keV) image of NGC 7130,
 adaptively smoothed 
and scaled
logarithmically.  
}
\end{figure}

In the total \chandra{} bandpass (0.3--8 keV),
the emission is
strongly concentrated around the nucleus, and also includes a
significant extended component.  
Figure \ref{fig:xrimg}a contains the broad-band image,
which has been adaptively smoothed using the CIAO task csmooth.
We employed the Gaussian kernel and smoothed 
to a minimum significance of 2$\sigma$ and a maximum scale of 5 pixels.

Using \chandra's simultaneous spectral and
spatial resolution, we show morphological differences within the 
X-ray bandpass.  The soft (0.3--1 keV) X-ray image (Figure \ref{fig:xrimg}b)
is very similar to the total image, with a strong central concentration
and significant extended emission that traces the galaxy's optical
structure.  Several strong individual sources outside
the nucleus are also evident.  In a medium X-ray band (1--2 keV), 
the significant emission is confined to a slightly smaller spatial
scale,
although several particular sources
or regions of enhanced emission outside the nucleus 
are still detected (Figure \ref{fig:xrimg}c).
In hard X-rays (2--8 keV), the largest-scale extended emission is 
absent (Figure \ref{fig:xrimg}d), but significant diffuse
emission is present, particularly north of the nucleus.
The nucleus is the strongest source in the image, 
and an additional source is also present. 
In the total and individual bands, 
the diffuse emission northwest of the nucleus is brighter 
than the extended emission toward the southeast.
In the soft image, for example, the diffuse emission has a
mean surface brightness of 2.2 and 1.3 counts arcsec$^{-2}$
in these two areas, respectively. 
This may be a consequence of extinction, with the
foreground inclined plane of NGC 7130 absorbing some of the
X-rays.

\begin{figure}
\includegraphics[width=0.5\textwidth]{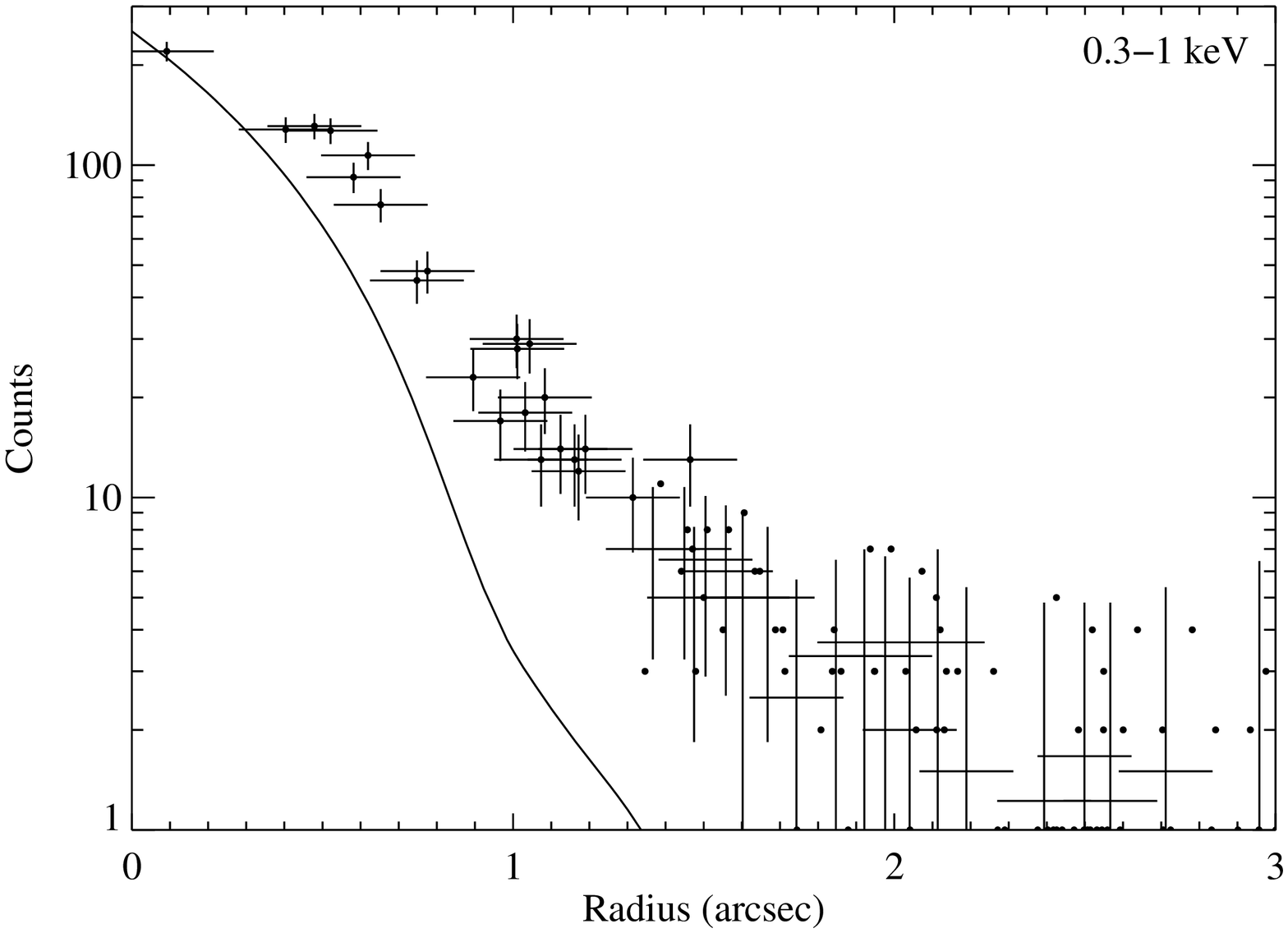} 

\includegraphics[width=0.5\textwidth]{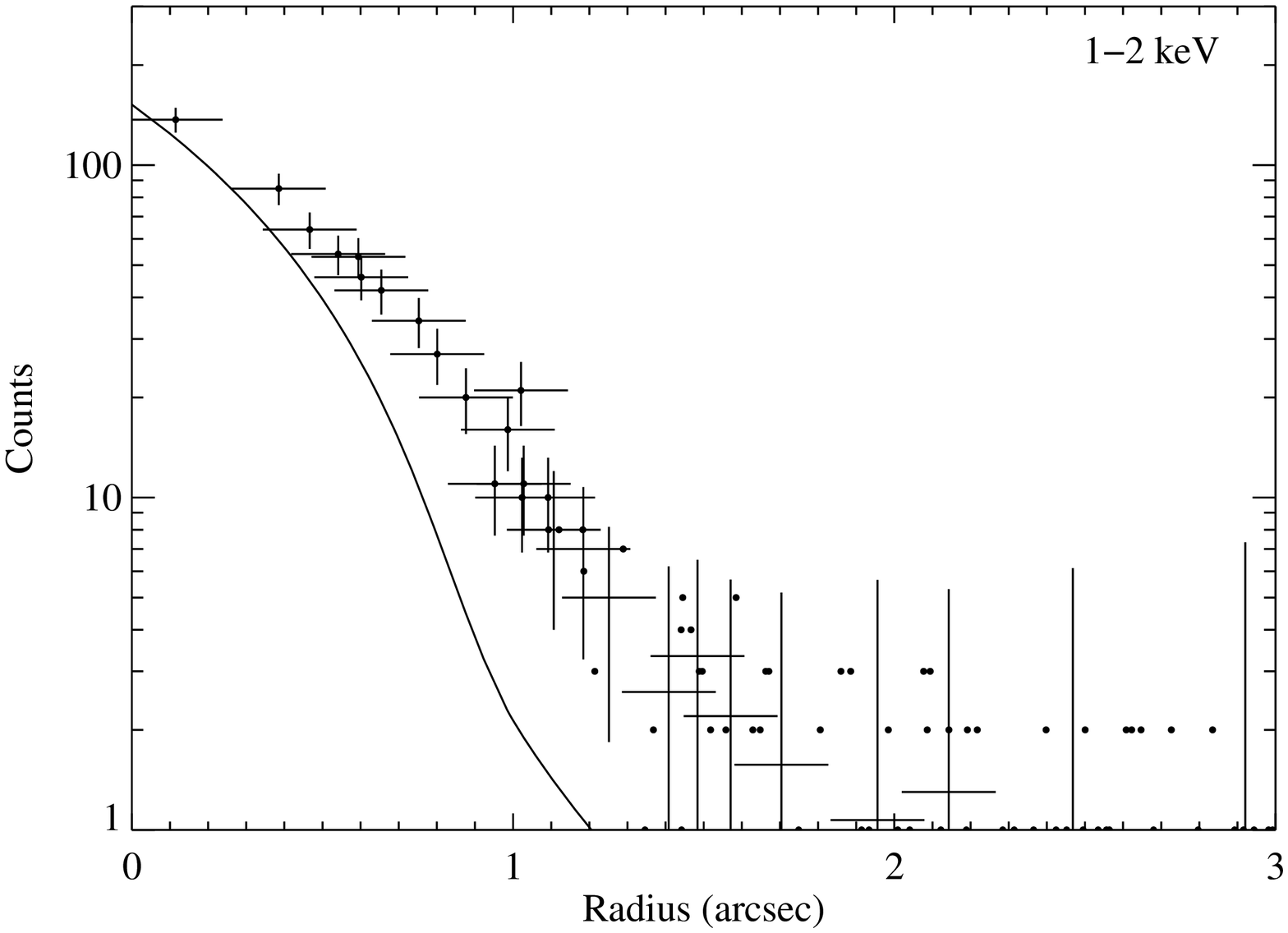} 

\includegraphics[width=0.5\textwidth]{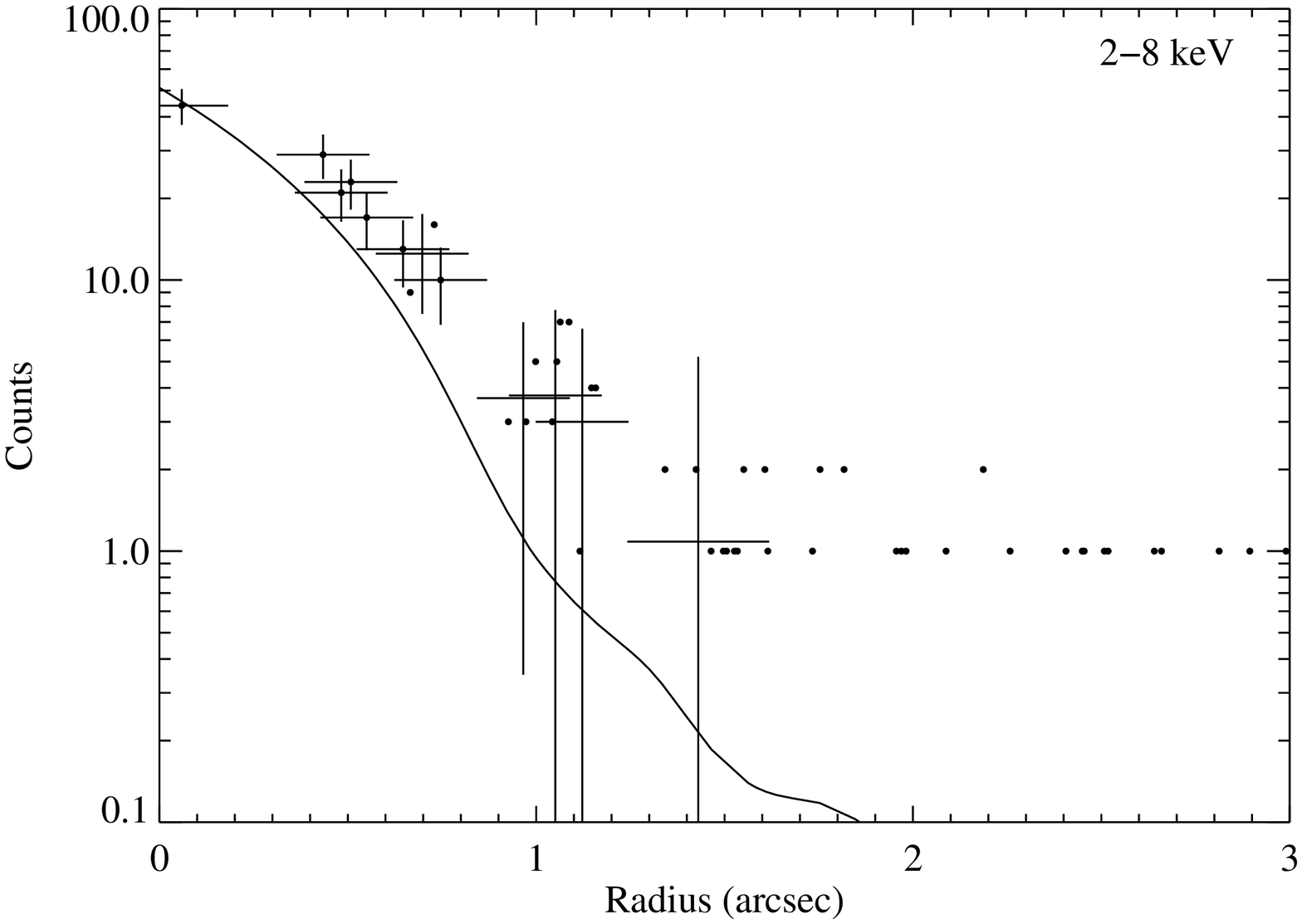} 
\caption{\label{fig:psf}
Radial profiles of the central emission at soft,
medium, and hard X-ray energies. 
Individual data points are marked with small filled circles.
Combined into bins that have
a minimum signal-to-noise of 3, 
these data and their errors are plotted as crosses.
In each case, the radial profile of the energy-appropriate PSF
scaled to the central peak is plotted as a solid line.
The small-scale emission is significantly resolved in the soft
and medium bands.
}
\end{figure}

In all three energy bands, emission is certainly extended on
large ($> 5\arcsec$) scales.
We compare the central source with \chandra's point-spread function
(PSF) in the three separate bands to determine its extent on smaller
scales.   If the AGN
were the sole source of nuclear emission, we would
expect the emission to be unresolved,
for the PSF FWHM $\approx 0\farcs8$, which corresponds to 270 pc at
the distance of NGC 7130.
In each of the three bands, we measure the  radial profile
of the observed emission and compare it with the radial profile of the 
PSF at the  appropriate energy
and detector position modeled with
the CIAO task mkpsf (Figure \ref{fig:psf}).
In each of the data and PSF images, we construct a radial profile by
associating the counts in an individual pixel with the pixel's
radial distance from the central emission peak.  
Thus, in each case, several data points are located within fractional
pixel distances of the center, within the FWHM of the PSF.
A two-dimensional Gaussian fit determines 
the location of the central peak, but it does not accurately 
characterize the shape of the PSF.
The individual data points measured in NGC 7130 are plotted
with small filled circles.  These data combined into bins of
signal-to-noise $\ge 3$ and their errors are plotted as crosses.
A smooth line connects the individual points measured in the
PSF images.

In the soft and medium bands, the observed radial
profile is significantly broader than the PSF on scales
smaller than 1\arcsec.
In the hard band, where the PSF is slightly broader,
the small-scale emission appears marginally resolved,
but the low signal-to-noise ratio precludes a significant measurement.
We conclude that the central source is
resolved at least at soft and medium energies,
so these broad-band emission peaks are not
due to the unresolvable AGN alone.

\begin{figure}
\includegraphics[width=6in]{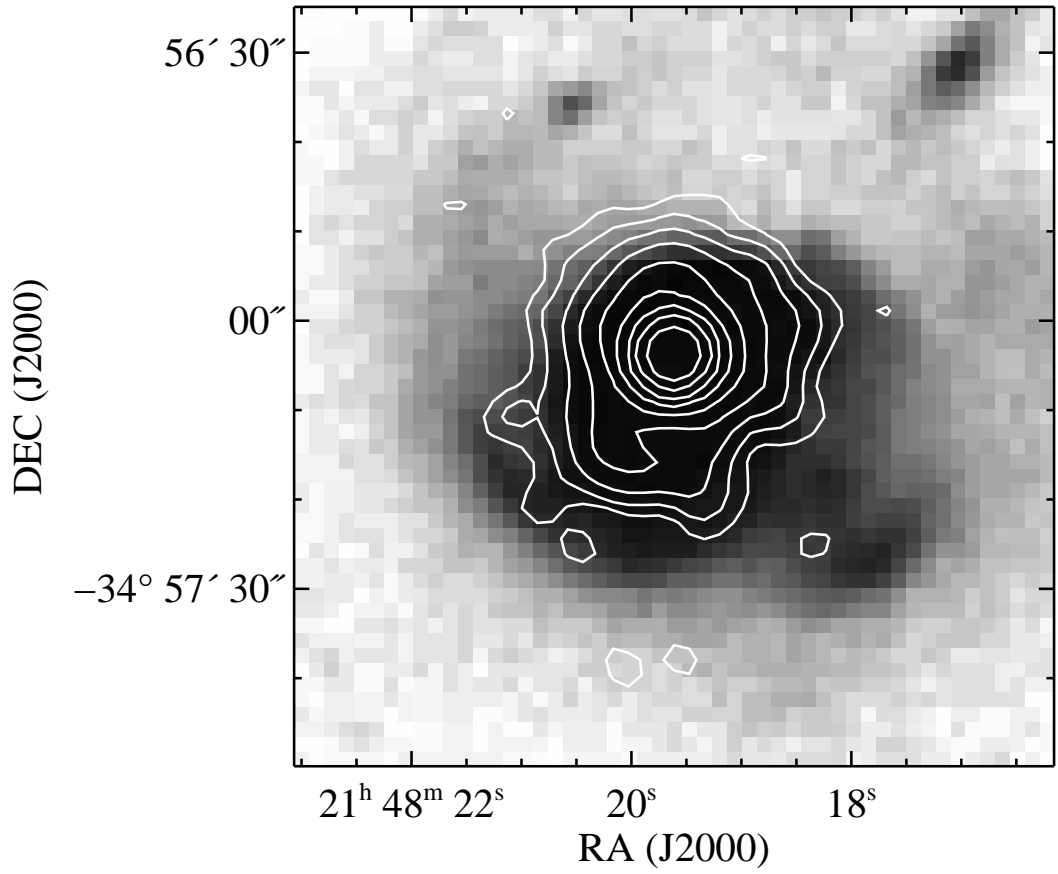} 
\caption{\label{fig:dss} 
Digitized Sky Survey 
image with total \chandra{} contours overlaid.
The X-ray data have been smoothed by a Gaussian of  FWHM = 5\arcsec{} 
to illustrate the full extent of the high-energy emission,
which covers most of the optically-bright galaxy.
}
\end{figure}

\begin{figure}
\includegraphics[width=6in]{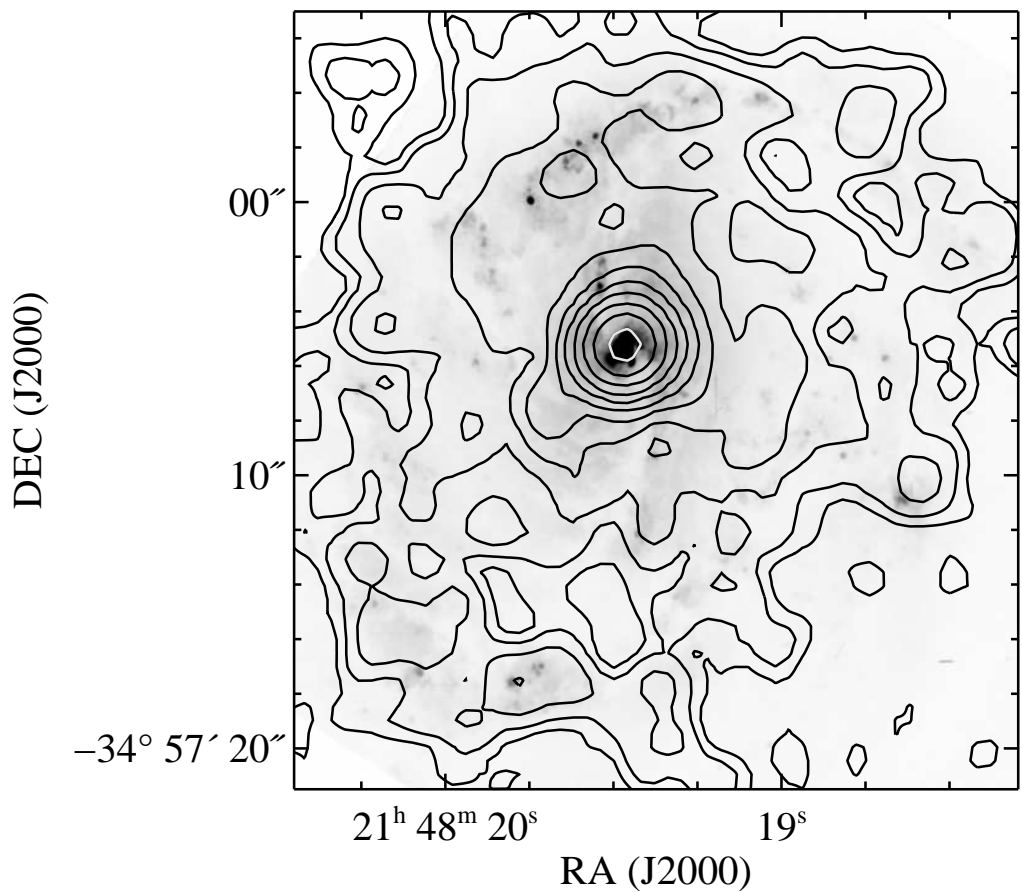} 
\caption{\label{fig:hst}
{\it HST} optical (6030\AA)  
image with \chandra{} total-band
contours overlaid. 
The high-resolution optical image shows  extended stellar
emission correlated with X-ray emission.
Well outside the nuclear region, on kpc scales, 
particular sites of intense star formation
are often also local X-ray enhancements.
The optical image is scaled linearly, and the X-ray
contours are scaled logarithmically 
by factors of 2 from 3$\sigma$ above
the background.
}
\end{figure}

As we demonstrate spectroscopically,
the nuclear emission is due to both the AGN and circumnuclear star formation,
while the extended emission is due to stellar processes alone.
Uniform smoothing of the 
data indicate the largest scale 
of the total X-ray emission of 38 $\times$ 36\arcsec{} 
(13 $\times$ 12 kpc).  The X-rays cover nearly the full extent of 
the galaxy's optical emission, as the X-ray contours overlaid on
the Digitized Sky Survey image illustrate (Figure \ref{fig:dss}).
On smaller scales the X-ray and optical emission are correlated.
Several low-level X-ray peaks located more than a kpc from the nucleus
are also sites of significant star formation, evident in 
an optical image obtained with the
{\it Hubble Space Telescope} ({\it HST}; Figure \ref{fig:hst}).
NGC 7130 was observed through the F606W filter using the Wide-Field and
Planetary Camera 2 for 500 s on 1994 August 23 (data set U2E60W01T).  
We use standard
processing of the optical image and shift the X-ray contours
$0\farcs75$ southwest to approximately align the bright nuclear
emission and other sources that appear at both wavelengths.
This shift is typical of the expected uncertainties in the astrometry
of the two observations.

\section{Spectroscopy\label{sec:spec}}
We extracted spectra of the nucleus and bright extended regions separately,
and we also considered a very extended region to measure the
X-ray emission of the galaxy as a whole.
In all cases, we measured the local background in surrounding
source-free regions and subtracted it.
Over the course of the \chandra{} mission, 
the soft X-ray sensitivity has diminished, likely the
result of build-up of material on the detector.
We have used the
``ACISABS'' model of G. Chatras and 
K. Getman\footnote{http://www.astro.psu.edu/users/chartas/xcontdir/xcont.html}
to create ancillary response files that account for
this time-varying effect.
We performed the model fitting in XSPEC \citep{Arn96}.
In the best-fitting models discussed below,
only data from 0.4 to 8.0 keV are considered, and 
the inclusion of additional model components and their
free parameters are significant at a minimum of the 95\% confidence
limit, based on an $F$ test.  All quoted errors are 90\% confidence
for one parameter of interest.

In all cases, the spectral models we consider here
begin with contributions typical of both AGNs and starburst galaxies.
The intrinsic AGN continuum observed in Seyfert 1 galaxies has
a photon index $\Gamma \approx 1.9$ \citep{Nan94}.
In Seyfert 2 galaxies, photoelectric absorption diminishes
the soft X-ray emission, although the intrinsic continuum can
be recovered at higher energies when 
the obscuration along the line of sight remains below $10^{24} \psc$.
In Compton thick AGNs such as NGC 7130, however, the
emergent continuum is entirely reflected and is strongly reprocessed,
to $\Gamma \approx 0$ \citep[e.g., ][]{Kro94} within the observed energy
regime.  
The Fe K$\alpha$ fluorescence line appears at
central energy $E_c = 6.4$ keV in material that is less
ionized than \ion{Fe}{17}, while He-like and H-like ions
produce $E_c = 6.7$ and 6.9 keV, respectively.
\citet{Ris99} suggested that NGC 7130 is
Compton thick based on the low ratio of 
2--10 keV/[\ion{O}{3}]$\lambda 5007$ flux
and indication of the Fe line in a low signal-to-noise 
spectrum obtained with {\it ASCA}.
The characteristic flat high-energy spectrum and prominent
(EW $> 1$ keV) Fe K$\alpha$ 
in the \chandra{} spectrum confirm this identification
\citep{Lev02}.

Most of the soft X-ray emission from a starburst is genuinely
diffuse thermal emission, with some portion due to  individual
point sources.
The diffuse emission, often in an galactic-scale outflowing wind
\citep{Hec90},
is spectrally soft and typically
requires multiple components at different temperatures.
We model these components with the ``MEKAL'' equilibrium plasma model 
in XSPEC \citep{Mew85,Mew86,Lie95}.
The integrated contribution of the unresolved
point sources produces a hard continuum that dominates
the starburst's X-ray emission above $\simeq$ 2 keV.
Surveys of starburst galaxies 
provide empirical support for these spectral models 
\citep{Dah98,Str04}.
Detailed studies of the nearest examples also 
explicitly separate the point source and diffuse contributions
\citep{Gri00,Wea02}.
Theoretically-motivated models of X-ray emission
from starbursts contain these general features
and further distinguish the continuum spectra of various
source populations \citep{Per02}, although we cannot
spectrally identify all the components of the synthetic spectra here.

\subsection{The Nuclear Region}
\begin{figure}
\includegraphics[angle=270,width=4.5in]{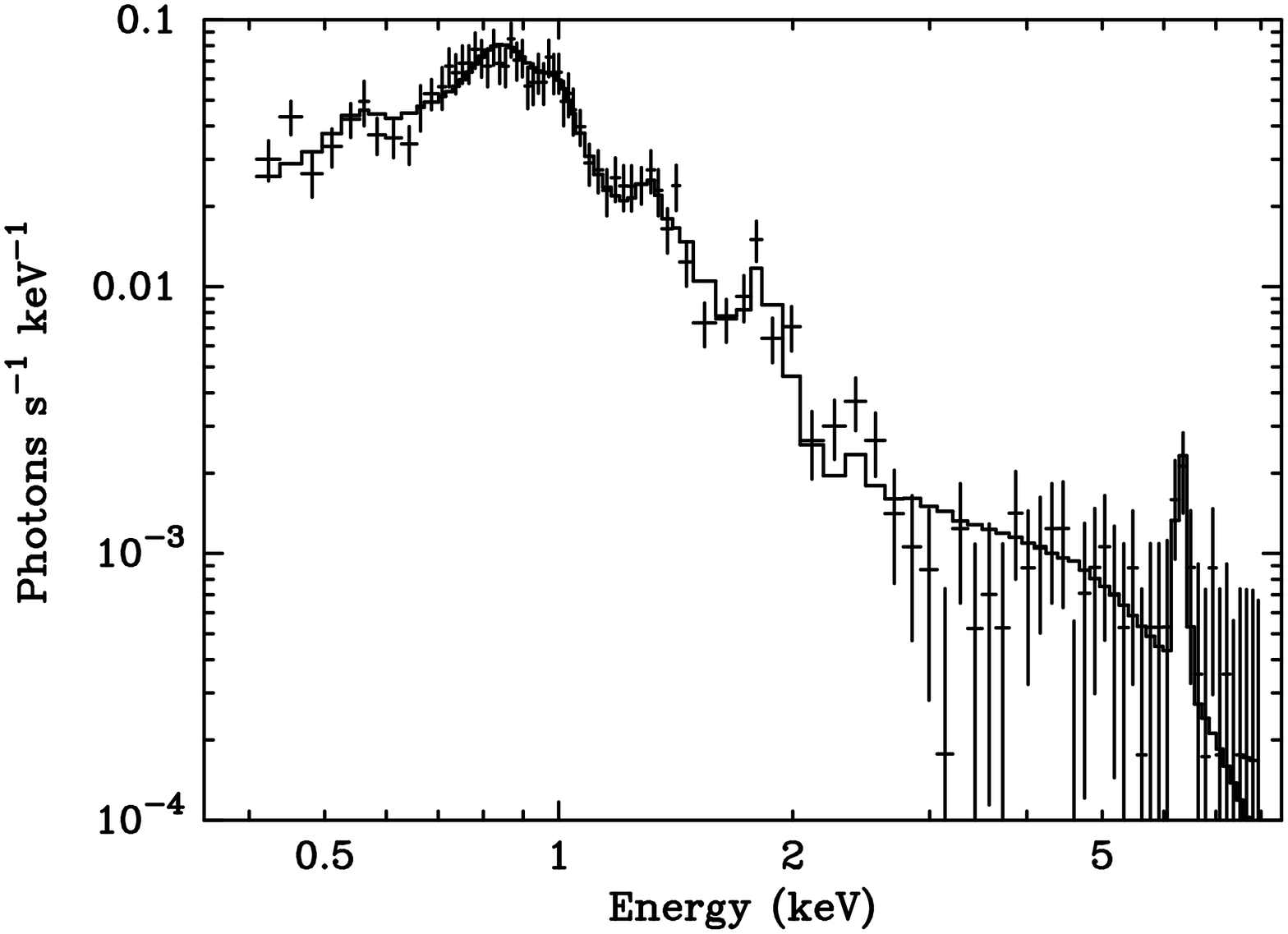} 
\includegraphics[angle=270,width=4.5in]{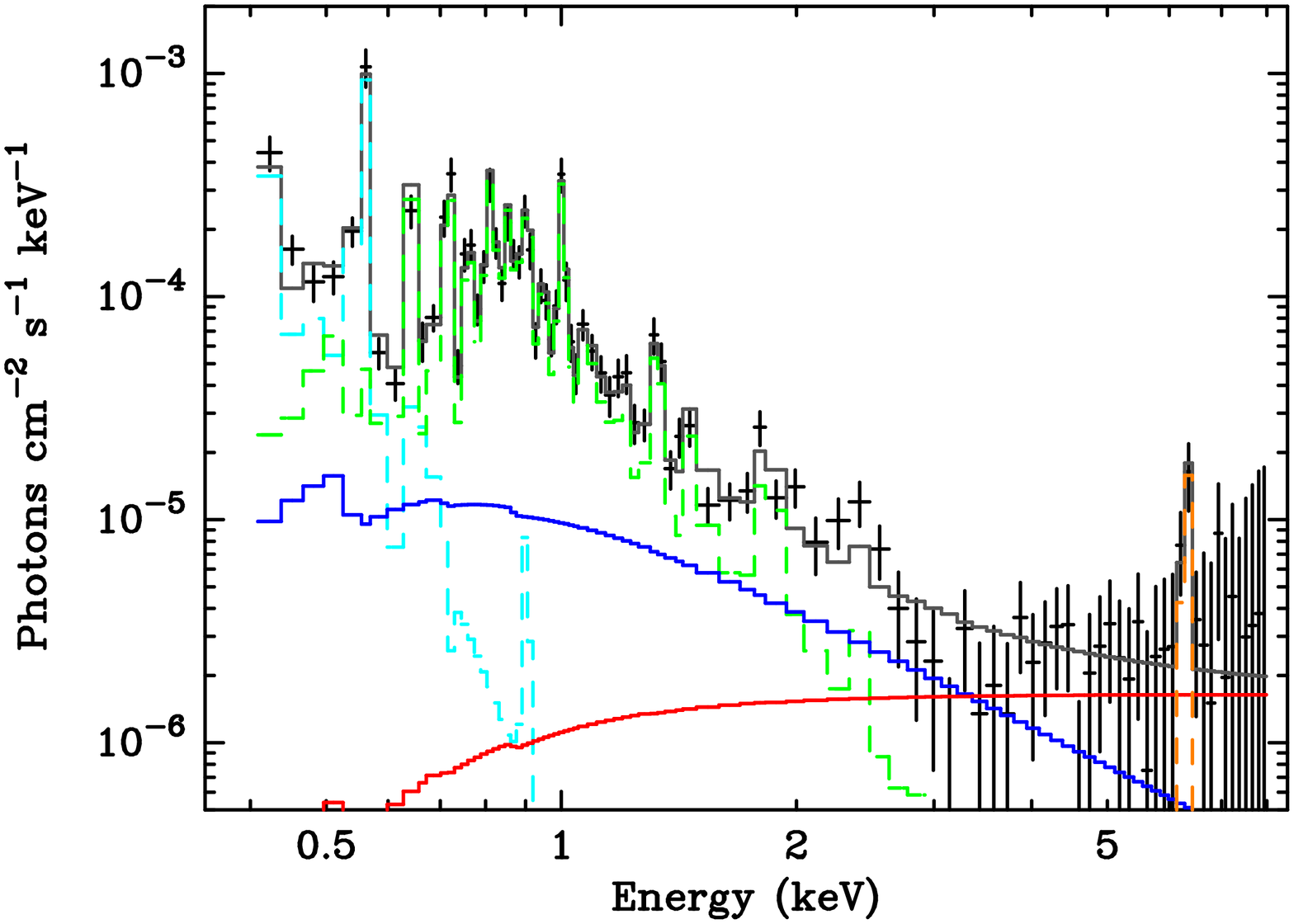} 
\caption{\label{fig:specnuc}
\chandra{} spectra of the nucleus of NGC 7130,
encompassing a region of  radius ${ 1\farcs5} \equiv 500 {\rm \, pc}$.  
The data have been rebinned
so each point represents a detection that is significant at
the 5$\sigma$ level, although the original unbinned data are
used in the model fitting.
({\it upper panel}) Observed data (crosses) and total model
convolved with detector response (histogram).
({\it lower panel}) Crosses show the intrinsic spectrum for the given model.
The total model (gray histogram) comprises
two thermal components (green and cyan) and 
a power law (blue) of the starburst, as well
as the reprocessed flat power law (red)
and Fe K$\alpha$ line (orange) of the AGN.
}
\end{figure}

The nuclear
aperture has a radius of ${ 1\farcs5}$.  
On this physical scale (500 pc), the nuclear spectral region encompasses
the compact starburst, as well as the AGN.   Thus, all the
physical components noted above are relevant.
In order to retain significant counts at energies above the Fe line,
we do not bin the data and use the C-statistic \citep{Cas79}.
The spectrum has a total of 2189 counts, and 
the best-fitting model includes two thermal components,
two power laws, and a Gaussian emission line.
All the emission is intrinsically absorbed in excess of the 
Milky Way column density ($N_{H,MW} = 1.7\times 10^{20}$; \citealt{Schl98})
along the line of sight to NGC 7130.
Figure \ref{fig:specnuc} shows the model fit to the data, and
the model parameters are listed in Table \ref{tab:modelpars},
with the total flux and statistical quality on the first line.
The components associated with the AGN and starburst are listed
separately on subsequent lines, although 
all parameters were fit simultaneously in the nuclear spectrum.

\subsubsection{AGN}
None of the intrinsic AGN emission emerges directly
from this Compton thick source.  The AGN is detected
only indirectly in the flat ($\Gamma = 0$) power law and the Fe line.
The  power law represents the reflected continuum.  
The hard X-rays that do emerge are not only diminished
in magnitude, but they are also strongly reprocessed.
Hence, we measure a very flat spectral index of this
power law component that contributes to the hard X-ray flux.
The buried AGN does not produce significant soft X-ray emission,
even within the nuclear aperture.

In this case, we have not applied more complete physical models,
such as those of \citet{Mag95}, which explicitly account for
the reprocessing of the intrinsic AGN continuum.  
Because all of the observable emission is reprocessed,
we have no sensitivity to the underlying physical conditions 
that determine the emission from this model, 
such as intrinsic continuum slope and viewing angle.
In fact, the only formal difference between the  flat power law
we employ and the reflection model is the Fe edge near 7 keV
of the latter, but with \chandra's diminishing sensitivity at
higher energies, we cannot measure this feature in the data.

The emission line at central energy 6.4 keV is consistent with
neutral iron fluorescence.  The large EW requires that 
the fluorescing region does not share our obscured view of
the AGN continuum.  
More exactly,  this prominent emission line
absolutely rules out any model in which the AGN
continuum is viewed directly through
$N_H < 10^{24}\psc$.
In addition to requiring a large
column density along the line of sight, 
the large EW constrains 
the covering fraction of this material.  
With best-fitting EW = 1.8 keV,  we expect 
the toroidal geometry of the obscuring region 
to have an opening angle less than about $30^\circ$, or
equivalently, a covering fraction greater than 75\%
\citep{Kro94,Lev02}.
Considering the uncertainty in the EW, however,
with a lower limit of 1 keV, the lower limit
on the covering fraction is 35\%.

\subsubsection{Starburst}
The weakly-absorbed power law and the
two thermal components are due to the circumnuclear starburst,
which dominates the soft X-ray emission of this region.
The data do not constrain the photon index of the power law well,
so we adopt the fixed value $\Gamma = 1.8$, reasonable for
the average of the X-ray binaries that produce the
non-thermal X-rays of starburst galaxies. 
The warmer thermal component
provides most of the photons, and its temperature
($kT = 0.6$ keV) is similar 
to that measured in  the inner regions 
of starbursts \citep{Str04}.
In NGC 7130, the very soft ($E < 0.8$ keV) emission is extremely strong,
which we model 
with the lower-temperature ($kT = 0.1$ keV) collisional plasma.
While this temperature is lower than the average observed
in  extended  starburst winds, multi-phase models of some starburst
disks, such as NGC 253, do include such a cool component \citep{Str02}.
In a starburst galaxy, 
the outflowing wind of hot gas is roughly conical, extending
from the compact base of star formation out of the
plane of the galaxy.
Most of the flux in the nuclear aperture
originates at the compact, bright base of the wind.
We appear to view this outflow  from an intermediate angle,
given both the  inclination of 
NGC 7130  \citep[$29^\circ$;][]{Whi92}, 
and the large-scale asymmetry of the diffuse soft X-ray emission.

We emphasize that {\em the bulk of the soft emission in the nuclear region
is not directly attributable to the AGN}.  Photoionization models
completely fail to reproduce the observed broad emission complexes
of the spectrum, especially
those around 1 keV, which
are characteristic of collisional excitation.
Existing models do not completely match the observed 
photoionized spectrum of NGC 1068 \citep{Kin02}, but the 
empirical spectrum of this galaxy
is also inadequate as a template for the soft X-ray emission of
NGC 7130. 
The strong thermal contribution is typical of starburst galaxies
and is therefore not surprising to find within the central 
500 pc of NGC 7130, which
definitely contains a powerful starburst.
Although NGC 7130 also hosts an active nucleus, its accretion is not the
only source of X-rays.  

X-ray spectra of ordinary starburst galaxies often
show non-solar abundances, especially enhanced abundances
of alpha elements relative to iron \citep{Mar02,Str04}.
Similar to the starburst galaxies that lack AGN,
we find enhanced abundances of alpha elements relative
to iron.
The best-fitting model of the nuclear region 
includes variable Fe abundance in the primary
(warmer) thermal component.  We 
find
 $Z_{Fe} = 0.38 (+0.1, -0.09) Z_\sun$, with 
(Fe/H)$_\sun = 4.68\times 10^{-5}$ \citep{And89},
fixing all other abundances at their solar values.
Alternatively, fixing Fe at solar abundance and allowing the
common abundance of O, Ne, Mg, Si, and Ca to vary, we
find $Z_{\alpha} = 2.4 (+0.9, -0.6) Z_\sun$.
In this case, the temperature is unchanged ($kT = 0.6$ keV),
although the spectral fit is slightly worse. 
The strongest of these lines 
are oxygen, so its abundance
dominates the determination of $Z_\alpha$, 
with (O/H)$_\sun = 8.51\times 10^{-4}$ \citep{And89}.
These results indicate that  Type II supernovae enrich
the thermal plasma, preferentially contributing $\alpha$ 
elements,  although
the measured $\alpha$-to-iron ratio is still lower than 
models predict for pure supernova ejecta 
\citep[see][and references therein]{Gib97}.

The small nuclear aperture isolates the AGN emission, but it
excludes some of the more extended, yet strong, central emission.
To better measure the total flux of the extended 
central starburst, we extracted the spectrum over a $3\farcs5$ 
($\equiv 1.1$ kpc) radius.  
As expected, the only significant
change is an increase in the total flux, to
$1.9 \times 10^{-13}$ and 
$2.2 \times 10^{-13} {\rm \,erg\,cm^{-2}\,s^{-1}}$
in the soft (0.5--2 keV) and hard (2--10 keV) bands, respectively. 
These values represent increases over the small nuclear
region of 15 and 22\% of the starburst flux in these two bands. 
The sense of the abundance variations in 
the more extended central emission is the same as
in the smaller region. 
We find 
$Z_{Fe} = 0.52 (\pm 0.1) Z_\sun$ for $Z_\alpha = Z_\sun$,
or
$Z_{\alpha} = 3.2 (+0.8, -0.6) Z_\sun$ for $Z_{Fe} = Z_\sun$.

Alternatively, we can spatially isolate the extended emission immediately
outside the nucleus in an annular aperture between 1.5 and $3\farcs5$ radii.
The extracted spectrum is soft, with no emission above 3 keV.
The starburst model, with one thermal component [$kT = 0.54$ (+0.08, -0.1) keV]
and a power law having fixed $\Gamma = 1.8$, fits well ($\chi^2/$dof$ = 12/10$).
The column density is not constrained, so
we fix it at the Galactic value.
The observed soft and hard fluxes are 
2.2 and  $1.7 \times10^{14} {\rm\, cm^{-2}\, s^{-1}}$, respectively.
The model underpredicts some very soft ($E <0.6$ keV) emission,
but an additional thermal component with $kT = 0.08$ keV is not
statistically significant. 
In this limited region, which includes only 285 counts,
no abundance variations are significant.

\subsection{Diffuse Emission}
\begin{figure}
\includegraphics[angle=270,width=4.5in]{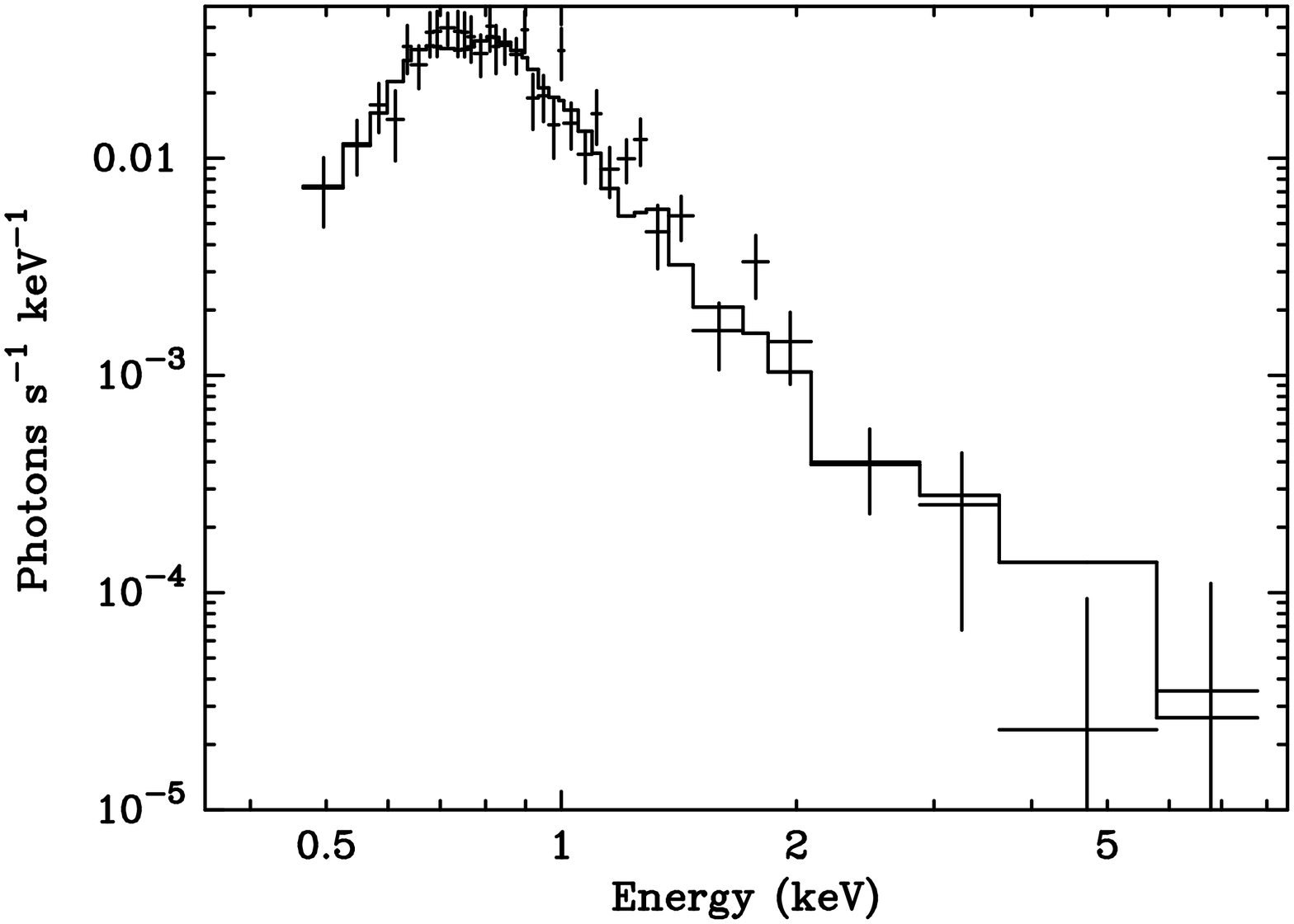} 
\includegraphics[angle=270,width=4.5in]{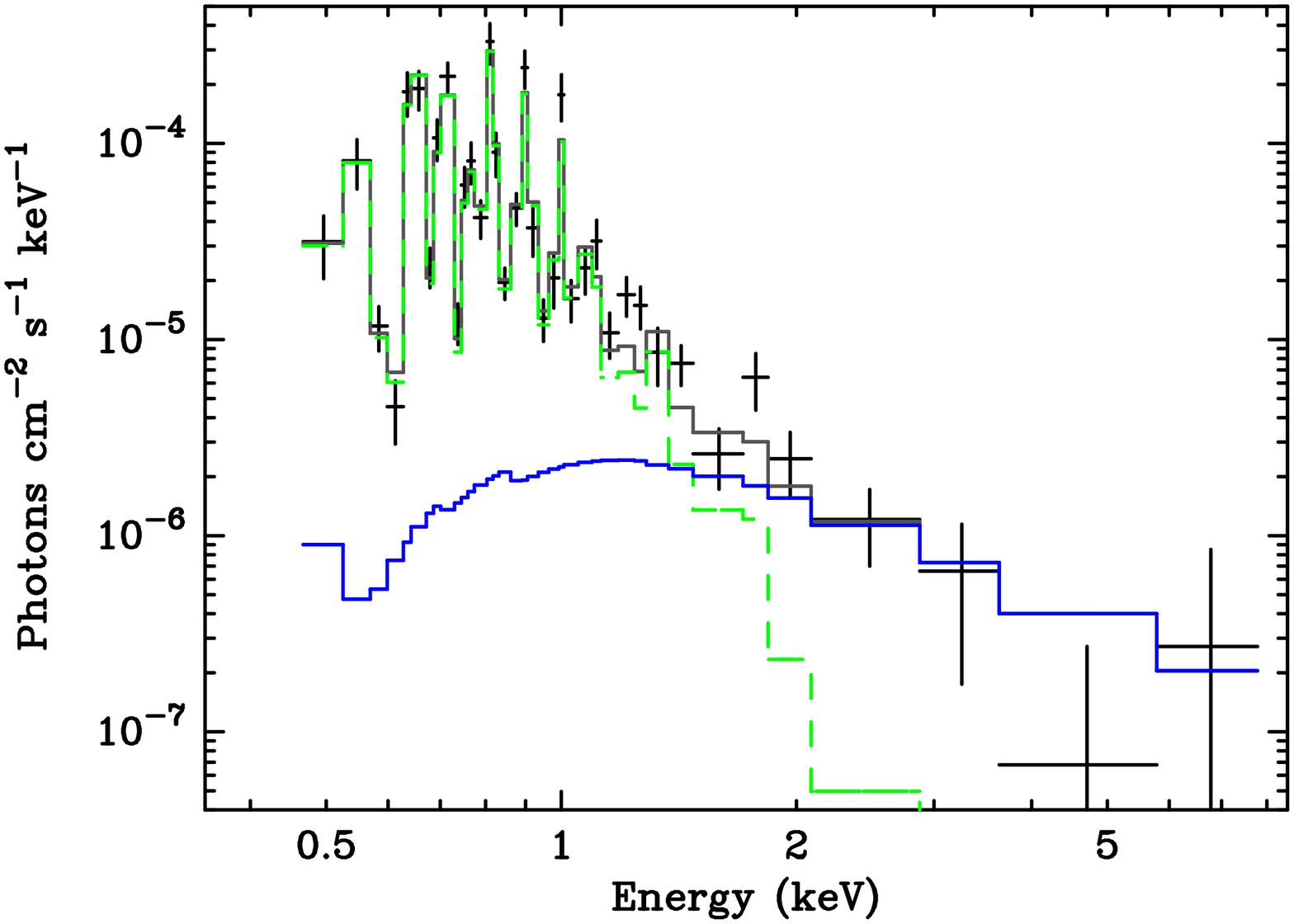} 
\caption{\label{fig:specx}
\chandra{} spectra of the diffuse, extra-nuclear
emission of NGC 7130.
({\it upper panel}) Observed data (crosses) and total model
convolved with detector response (histogram).
({\it lower panel}) 
Crosses show the intrinsic spectrum for the given model.
The gray histogram is the total model, which requires
a thermal component (green) and a hard continuum (blue).
}
\end{figure}

We isolated the diffuse emission outside the nucleus
in an annulus of outer radius $15\arcsec$ and
inner radius $4\arcsec$,  
and also excluding one bright source.
This spectrum is extremely soft and requires only 
two of the starburst components of the
nuclear model: a thermal contribution and a power law.
The AGN components of the flat power law and the fluorescent line are absent.
Allowing the column density to be a free parameter, we find that 
the obscuration
would be much greater than in the nuclear region 
[$N_H = 4.3 (+0.1, -0.3) \times 10^{21}$], so we fix it at the 
Galactic value.
The spectrum and model fit are shown in Figure \ref{fig:specx}, and
the best-fitting parameters of this model are listed in Table \ref{tab:modelpars}.
The temperature here is intermediate between the two temperatures of
the nuclear region.
Although the spectrum is soft, the non-thermal continuum
is essential and accounts for the detectable hard emission.
The significant non-thermal continuum in this extra-nuclear
region thus
supports the interpretation of the steeper nuclear power
law as a starburst component.

The surface brightness of the diffuse emission is much lower than that
of the nuclear region, even considering only the starburst-associated
portion.  Extending over a much larger area, however, the
diffuse emission contributes substantially to the total stellar
luminosity.  In both soft and hard bands,
the  luminosity of the diffuse emission is
30\% of the nuclear starburst luminosity.

\subsection{Additional Sources}
We detected several additional sources in NGC 7130.
The sources are detected with non-zero
spatial extent using the CIAO wavdetect algorithm.  
We considered total (0.3--8 keV), low-energy (0.3--2 keV)
and high-energy (2--8 keV) images separately, but the sources
of the latter two images are a subset of sources in the
total-band image.
In all cases, we measure the counts in a 2\arcsec{} 
radius aperture after subtracting
a local background.  These source locations
and their net counts in total,
soft, and hard energy bands are listed
in columns 2 through 6, respectively, of Table \ref{tab:ptsrc}.
Most of the sources are extremely soft, with nearly all
counts in the soft band and specifically below 1 keV.  
Two exceptions are sources
3 and 4.  Source 3 is evident in the 1--2 keV image (Figure \ref{fig:xrimg}c)
north of the nucleus.
Source 4 is located southeast of the nucleus, most prominent in
the hard X-ray image (Figure \ref{fig:xrimg}d), but also
obvious in Figure \ref{fig:xrimg}c.

To estimate fluxes and luminosities, we 
model each spectrum as an absorbed power law with 
fixed $\Gamma =1.8$ and free column density in excess of
the Milky Way absorption along the line of sight.
While this may not be a realistic model for these sources,
the procedure yields consistent results for the observed fluxes
and luminosities, which we list
in columns 7 and 8 of Table \ref{tab:ptsrc}.
To minimize errors due to the model-dependent estimate of fluxes
and luminosities, we quote these values in the 0.3--8 keV band
and do not correct the luminosity for absorption.

Each source measurement covers a region of 650 pc scale
and could include clusters of stars;
these are not necessarily individual X-ray sources. 
The more luminous ($L> 10^{39} {\rm \, erg \, s^{-1}}$)
sources are intriguing, however, 
because they exceed the Eddington luminosity of a stellar-mass
black hole.  If they are individual sources, they could
accrete at a higher rate, emit anisotropically, or
be more massive than a single neutron star.
(See \citealt{Mil04} and \citealt{van04} for reviews.)

\section{Luminosity Contributions and Spectral Energy Distributions\label{sec:sed}} 
We list various contributions to the  observed luminosity in
Table \ref{tab:lum}.  To calculate the  luminosity of NGC 7130
as a whole, we have examined the  spectrum of the  
$38\arcsec$ (13 kpc)   
emission extent.  This region obviously
encompasses a range of physically distinct emission sources
but provides a reasonable estimate of the total flux. 
The AGN and starburst components of the bright central region
remain strongest, with the temperature of the dominant thermal
component slightly lower (0.56 keV) than its counterpart in the
nuclear spectrum.
The AGN contribution listed in the table 
includes only the flat power law and Fe line
measured in the small nuclear aperture.  The 
central starburst
includes the thermal emission and  power law of the larger 
($3\farcs5$) nuclear aperture.  The extended starburst includes only the
emission outside the nucleus yet within the central $30\arcsec$,
excluding the bright source 4.
The total galactic emission exceeds the sum of individual regions
because this aperture covers a larger area.
The gaps between the individual interior apertures serve to
isolate physically-distinct regions, and individual sources are
excluded from the diffuse region.

In the integrated spectrum, the soft and hard X-ray emission are comparable,
yet they each have a distinct origin.
Nearly all of the detected soft X-rays are due to 
the central starburst, 
with the AGN directly accounting for 
almost none (1\%)  of the total soft emission.
The AGN is responsible for most (60\%) of the detected
hard X-rays, though 
an appreciable fraction (40\%) of the total hard flux is stellar.

\begin{figure}
\includegraphics[width=6in]{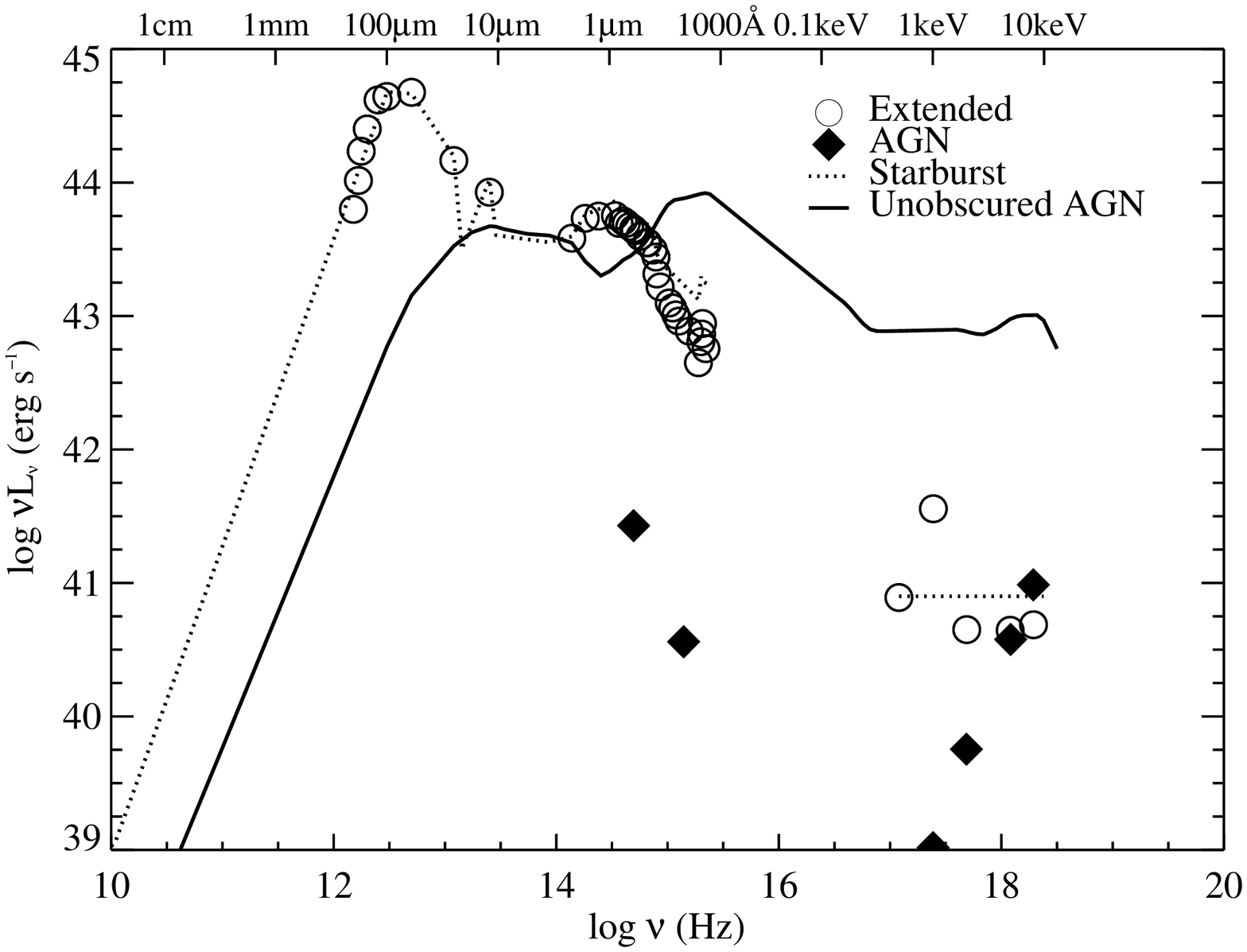} 
\caption{\label{fig:sed}
Spectral energy distributions of NGC 7130,
 showing large-scale emission
(open circles) and the spatially and spectrally
isolated AGN  contributions (filled diamonds).
On large scales, the galaxy appears very similar to dusty
starbursts (dotted line), from \citet{Schm97}, with the 
X-ray scaling of  \citet{Ran03}.  
For comparison, the average radio-quiet quasar spectrum \citep{Elv94} scaled to
the estimated intrinsic bolometric luminosity of NGC 7130 is also plotted 
(solid line).
}
\end{figure}

The relative contributions of the AGN and  star 
formation are also evident in spectral energy distributions
(SEDs) of NGC 7130.
On large scales, the observed emission of NGC 7130 is very similar
to ordinary starburst galaxies at most energies.
In Figure \ref{fig:sed}, 
a rough scaling of  the median reddened starburst, which
\citet{Schm97} determined empirically, passes through most of
measurements of NGC 7130 on scales larger than 2\arcsec.
The starburst SED is extended to the X-rays using the
relationships of \citet*{Ran03}.
The published data are from \citet{Spi02}, the {\it IRAS}
Faint Source Catalog \citep{Mos90},
the 2MASS Extended Object Catalog, \citet{Sto95}, and \citet{Kin93}.
The extended X-ray measurements are integrated over
the full 13 kpc galactic diameter and 
{\em exclude} the spectrally-identified
AGN components.  
Even at the low resolution plotted in the SED, the strong
thermal contribution at 1 keV is obvious.

The majority of the luminosity emerges at far-infrared (FIR) wavelengths,
and the spectral shape of this regime is characteristic
of star-forming galaxies.  While active galaxies
have typical 25- to 60-$\mu$m flux ratios
$f_{25}/f_{60} > 0.26$ \citep{deG87},
in NGC 7130 $f_{25}/f_{60} = 0.13$. 
The FIR luminosity $L_{FIR} = 4.9\times10^{44} {\rm \, erg\, s^{-1}}$,
and the total 8--1000$\mu$m luminosity
$L_{IR} = 9.0\times10^{44} {\rm \, erg\, s^{-1}}$, following
the prescriptions of \citet{Hel85} and \citet{San96} to calculate
these luminosities from {\it IRAS} fluxes.
In total, the large-scale emission of NGC 7130 is primarily stellar.

Overall, the relationship between FIR and X-ray emission is typical of
star-forming galaxies, and they yield consistent estimates of
the star-formation rate (SFR).  
In starburst galaxies, $L_{bol} \approx L_{IR}$, 
so for $SFR =  L_{bol}/2.8\times10^{43}$ \citep{Lei95}, we find
 $SFR = 32 M_{\sun}\, {\rm yr^{-1}}$.
This total SFR is consistent with the empirical X-ray scaling 
of \citet{Ran03}, who find $SFR =  L_{0.5-2}/4.5\times10^{39}$
in terms of the 0.5--2 keV luminosity.
This X-ray correlation  yields $SFR = 34 M_{\sun}\, {\rm yr^{-1}}$ over 
NGC 7130 as a whole, where we exclude the AGN contribution to the
soft X-ray luminosity.
The spatial concentration of the starburst's X-ray flux
indicates that most of the star formation is centrally concentrated,
with 70\% occurring in the central 1.1 kpc region.

The correlation of the SFR with 2--10 keV luminosity
($SFR =  L_{2-10}/5.0\times10^{39}$; \citealt{Ran03}) 
yields  the lower value of 
$SFR = 12 M_{\sun}\, {\rm yr^{-1}}$, again excluding the AGN
contribution to $L_{2-10}$.  
As \citet{Per04} note, the luminosity of
young populations alone should reflect the current SFR.
In NGC 7130, any such ``correction'' to the starburst's hard X-ray luminosity
would further underestimate the SFR compared with the IR-determined value.
We conclude that contributions from
older populations (namely low-mass X-ray binaries)
to the stellar $L_{2-10}$ are minimal.
Instead, current star formation, including high-mass X-ray binaries
and ultra-luminous X-ray sources \citep{Col04}, dominate 
the hard X-ray luminosity of the NGC 7130 starburst.

We identify the specific contribution of the AGN at several wavelengths.
In X-rays, we isolate the AGN spectrally in the small nuclear aperture,
considering only the flat power-law continuum and fluorescent line emission
components.
In \hst{} images we spatially isolate the AGN.
In addition to the optical data described above, NGC 7130
was observed for 1,984 s with the Faint Object Camera at 2150\AA{}
(through the F210M filter; data set X2RN0701T), 
and for 320 s with 
NICMOS at 1.6$\mu$m (through the F160W filter; data set N3ZB25010). 
\citet{Gon98} identify the nucleus with a knot of faint UV emission.
This is a bright near-infrared (NIR) source, and 
two dust lanes converge at this location in both the optical and NIR images.
For consistency, we determine the nuclear flux the same way in all
three \hst{} images.
We measure the background-subtracted 
source flux in a small aperture to minimize contamination from
nearby sources, then correct for the flux that falls
outside the aperture using models of the 
point spread function\footnote{http://www.stsci.edu/software/tinytim}.
In the NIR, optical, and UV bands, we measure
unresolved flux densities of
$8.2\times10^{-4}$,
$9.5\times10^{-5}$, and
$4.5\times10^{-6}$ Jy,
respectively.

In total, the AGN contributes very little to the observed luminosity
from 1mm to 10 keV.  The agreement between the
FIR/X-ray ratio of starburst galaxies and 
the observed quantities demonstrates that 
the AGN cannot be 
dominant contribution to the bolometric luminosity, either,
because the reprocessed radiation would emerge in the FIR.
We roughly estimate the intrinsic luminosity of the active nucleus 
from the X-ray observations and models.  
The Fe K$\alpha$ luminosity is related to the 
intrinsic AGN 2--10 keV luminosity, $L_{2-10,AGN}$.
For the measured EW, we estimate
$L_{2-10,AGN} = 1\times10^{43} {\rm \, erg\, s^{-1}}$ from
Monte Carlo simulations of Compton thick reprocessing \citep{Kro94,Lev02}.
Scaling  by the empirical measurements of radio quiet quasars
\citep{Elv94}, we find 
for the bolometric luminosity of the AGN below 10 keV
$L_{bol,AGN} = 4\times10^{44} {\rm \, erg\, s^{-1}}$.
With the observed  $L_{IR} = 9.0\times10^{44} {\rm \, erg\, s^{-1}}$,
we conclude that the AGN provides less than half the bolometric
luminosity of NGC 7130.

In its total SED, NGC 7130 is
similar to other galaxies that contain a
Compton thick AGN along  with
a strong starburst.  
In some examples, 
such as
Arp 299 \citep{Del02} and NGC 4945 \citep{Iwa93,Don96}, 
the AGNs are identified exclusively 
at X-ray energies,
and the starbursts truly dominate the total emission.
Observations at energies
greater than 10 keV effectively reveal these AGNs, but
the bolometric luminosities of the active nuclei are
only one-tenth of the totals measured in the FIR.
For Arp 299, $L_{2-10,AGN} = 5.6\times10^{42} {\rm \, erg\,s^{-1}}$ 
\citep{Del02},
or  $L_{bol,AGN} = 1.7\times10^{44} {\rm \, erg\, s^{-1}}$
(at a distance of 47.5 Mpc for  $H_0=70 {\rm \, km\,s^{-1}\,Mpc^{-1}}$).
Using the integrated FIR flux densities of \citet{Soi89},
$L_{IR} = 2.9\times10^{45} {\rm \, erg\, s^{-1}}$. 
For NGC 4945, $L_{2-10,AGN} = 3\times10^{41} {\rm \, erg\,s^{-1}}$ 
\citep{Gua00}, 
$L_{bol,AGN} = 1.1\times10^{43} {\rm \, erg\, s^{-1}}$,
and 
$L_{IR} = 1.0\times10^{44} {\rm \, erg\, s^{-1}}$
\citep{Ric88}.
We note that the luminosities we quote for NGC 4945 differ
from those of \citet{Gua00}, who claim that the AGN emission
predominates. 
The most significant difference is that 
here we consistently use the 
3.7 Mpc distance to  NGC 4945 \citep{Mau96}, rather than the 
luminosity distance of 11 Mpc that \citet{Gua00}
apply in their X-ray calculations.

Spectroscopy over large areas
that is limited to $E< 10$ keV does not yield accurate measurements
of these AGNs in starburst-dominated galaxies.
\citet{Don03} demonstrate for NGC 4945 that most of the observed 2--10 keV
flux in the roughly 1 degree$^2$ apertures of the
{\it Rossi X-Ray  Timing Explorer} instruments
is {\em not} the nuclear emission that \chandra{} resolves.
Lack of spatial discrimination likely accounts for the
unusually small  Fe K$\alpha$ EW of Arp 299.
In an
{\it XMM-Newton} aperture of 8.6 kpc diameter,
\citet{Bal04} find EW = 422 eV in NGC 3690,
the host of the the heavily absorbed nucleus of 
the Arp 299 merger. 
High spatial resolution data from \chandra{}
show that only 15\% of the observed 2--10 keV flux
is due to this nucleus \citep{Zez03}, but
these data are not sensitive enough to determine 
the line EW against the nuclear continuum alone.
Instead,  measured against the combined 
continuum of an ensemble of sources,
the line appears to be very weak
compared with most Compton thick AGNs, where
EW $\ge 1$ keV is typical.

\section{Implications for Surveys\label{sec:impl}}

NGC 7130 is an instructive example of the consequences of 
star formation and Compton thick obscuration for 
X-ray surveys to identify AGN. 
When the active nucleus is the only source of X-ray emission,
the simple measure of a hardness ratio of just two energy bands
characterizes it well.
Especially when the obscuration is Compton thin and some 
of its intrinsic power is directly detected, 
a hardness ratio provides a
good estimate of both the intrinsic AGN luminosity and the
obscuring column density.  Compton thick obscuration and
coincident starbursts both complicate this simple scenario,
however, even in nearby examples.
Because of the stellar component, 
the total soft X-ray emission of NGC 7130 is very strong.
If the hardness ratio is calibrated against
an absorbed power law to model the X-ray emission of an AGN alone,
NGC 7130 would mistakenly appear to be completely  {\em unobscured},
and the intrinsic luminosity of the buried central engine
would  be severely underestimated.

\begin{figure}
\includegraphics[width=6in]{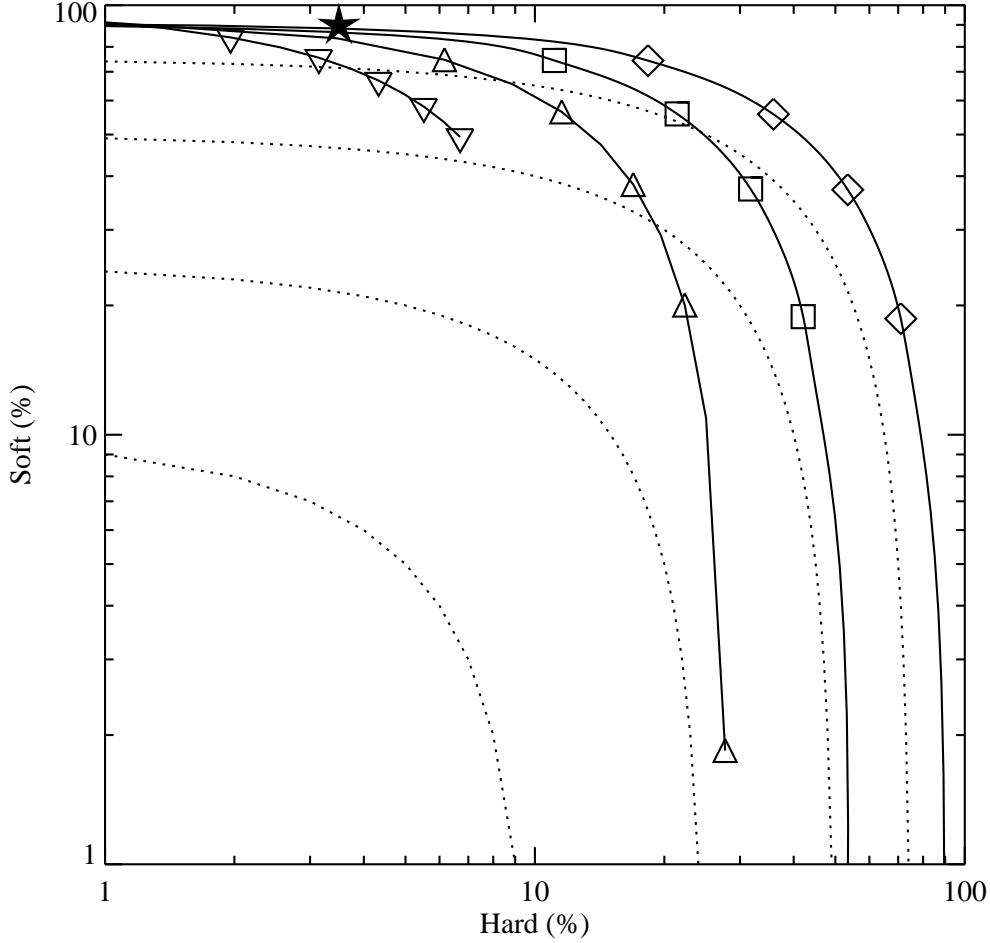} 
\caption{\label{fig:hardness}
Percentage of total observed counts in the
soft (0.3--2 keV) and
hard (5--8 keV) bands as a function of AGN obscuration and
starburst fraction. 
Along solid curves, AGN obscuration is fixed, at
$N_H = 10^{22}$ (inverted triangles),  $10^{23}$ (triangles),   
$3\times10^{23}$ (squares), and $10^{24}\psc$  (diamonds).
In each case, the symbols are plotted at
starburst contributions of 0, 20, 40, 60, and 80\%,
from lower right to upper left.
A pure starburst would appear at (hard, soft) = (0.8, 93).
Contours of constant medium-energy (2--5 keV) contribution are plotted with
dotted lines, for 90, 75, 50, and 25\% from lower left to upper right.
NGC 7130 is plotted as a filled star near (4, 90).
}
\end{figure}

We can use NGC 7130 to extend our analysis to larger surveys.
We develop a general diagnostic using X-ray colors to
distinguish between the starburst contribution and obscuration.
We define soft (0.3--2 keV), medium (2--5 keV), and hard (5--8 keV)
energy ranges. 
To identify the discriminating characteristics, 
we consider variations of  AGN obscuration and starburst strength
in a composite system.
In this case, the AGN is modeled as a single power law, with $\Gamma = 1.9$,
and column density ranging 
from $N_H = 10^{22}$ to $N_H = 10^{24}\psc$.
The starburst  is modeled after  the central region of NGC 7130,
with a strong 0.6-keV thermal plasma, a fainter soft component,
and a $\Gamma=1.8$ power-law continuum. 
We combine each AGN with a starburst whose strength
varies  from 0 to 100\% of the total counts.
We plot the results (Figure \ref{fig:hardness})
in terms of the percentage of hard and soft counts
\chandra{} would detect, so these spectral ratios
can be utilized without detailed  modeling.

In general, the  three broad X-ray bandpasses
distinguish among a range of starburst and AGN properties.
Some confusion remains in the combined cases of high
obscuration and large starburst fraction, where the
hard count rate is low.  No discrimination is possible 
when the fraction of hard
counts is less than about 2\%, which  occurs
for a  starburst fraction in excess of about 90\%.
With some significant hard X-ray emission, 
the AGN of NGC 7130 would at least be characterized as strongly obscured,
although  the X-ray ``colors'' alone  do not identify it certainly
as a Compton thick case.

At high redshift, however, low flux 
prevents even  X-ray identification of an active nucleus similar to that
of NGC 7130.
If NGC 7130 were located at $z=1$, 
the  0.5--2  and 2--10 keV observed fluxes would be
$3\times10^{-18}$ and $1\times10^{-17} {\rm \, erg\, cm^{-2}\, s^{-1}}$,
respectively.  
At $z=2$, the observed fluxes would be an order of magnitude lower.
Despite the benefit of the high-energy emission that is
redshifted into the observable bandpass, such
Compton thick Seyferts would be undetectable, 
even in the deepest current surveys.
While these surveys 
successfully resolve most of the X-ray background
below 10 keV \citep[e.g.,][]{Mus00,Bra01,Gia01,Has01}, 
their sources do not successfully reproduce the
spectrum at higher energies. 
 A population of Compton thick
AGN is essential \citep[e.g.,][]{Set89,Com95},
in addition to the X-ray sources that
are observed directly. 
The example of NGC 7130 demonstrates why this 
obscured class is
not evident from  X-ray measurements below 10 keV alone.

Although Compton thick 
AGN are missing from current X-ray surveys,
the existence of this population 
does not conflict with
current results on black hole demographics or 
multiwavelength background measurements.  The
host galaxies are observed and included in 
direct estimates of black hole density.
These sources 
may be evident
in the similar luminosity functions of 
low-luminosity AGN and star-forming galaxies.
The evolution of AGNs
is a function of luminosity, with the density
of lower-luminosity AGNs peaking at lower redshift
($z<1$; \citealt{Cow03,Ued03}), similar
to IR-bright starburst galaxies \citep{Cha01}.
The starburst-dominated examples, including 
Arp 299 and NGC 4945, explicitly demonstrate that
these two classes have some members in common.
Future, more sensitive missions, 
such as the {\it X-ray Evolving Universe Spectrometer},
with planned flux limits of $10^{-17} {\rm \, erg\, cm^{-2}\, s^{-1}}$, 
{\em will} be able to measure these obscured AGNs 
at their dominant epoch.

\section{Conclusions\label{sec:concl}}
Nearby Seyfert/starburst composite galaxies allow detailed examination
of high signal-to-noise observations.  Locally, the composite galaxies
are common rather than exceptional, and they serve as examples of the
complications that cannot feasibly be disentangled in more distant
sources.  In this case study of NGC 7130, we discern the multiple
physical processes that produce its intrinsic luminosity.  
Overall, star formation dominates the 0.3--8 keV
luminosity, with both a powerful
concentrated circumnuclear starburst and
an extended diffuse disk component together accounting for 70\% of
the detectable X-rays.
The AGN produces the majority of the  observed emission only at higher
energies, above 3 keV.
The effects
of obscuration and reprocessing vary with observed wavelengths and
determine the emergent, observable emission.  
The obscuration of the nucleus of NGC 7130 is
Compton thick, preventing direct detection of the intrinsic
emission in the \chandra{} bandpass, below 8 keV.
The stellar light is efficiently reprocessed to FIR wavelengths,
where it provides the  bulk of the bolometric luminosity.

The example of NGC
7130 empirically demonstrates that an optically-evident AGN can be
hidden or confused, even at X-ray energies.  
Integrating over large spatial scales, 
the properties of the AGN  would be misidentified,
although broad-band flux measurements
can be used to characterize the nucleus roughly.  
If its host galaxy were more distant, the active
nucleus would not be detected at all.  These buried AGNs are directly
relevant to the study of the cosmic X-ray background, and NGC 7130
illustrates the difficulty of finding them.

NGC 7130 and similar galaxies are also relevant to the
study of ultraluminous infrared galaxies.  
At issue in this class is the underlying energy source, which
may be an active nucleus or star formation. 
While the local composite galaxies are not luminous enough to
qualify as direct analogs of the more distant and powerful
ultraluminous galaxies, they illustrate the component building blocks that
may be present in the latter.
The investigation of NGC 7130 also suggests
methods to discriminate star formation from accretion activity,
with the conclusion that over large physical scales, star formation
can dominate even when a normal active nucleus is present.

\acknowledgements
We thank the referee, R. Della Ceca, for useful suggestions. 
This research has made use of the NASA/IPAC Extragalactic Database
(NED) which is operated by the Jet Propulsion Laboratory, California
Institute of Technology, under contract with the National Aeronautics
and Space Administration.
The Digitized Sky Survey was produced at the Space Telescope Science
Institute under U.S. Government grant NAG W-2166. The 
image presented here 
was made by the Royal Observatory Edinburgh with the UK Schmidt Telescope.
Part of this work was 
based on observations made with the NASA/ESA Hubble Space Telescope,
obtained from the data archive at the Space Telescope Science
Institute. STScI is operated by the Association of Universities for
Research in Astronomy, Inc. under NASA contract NAS 5-26555.
This publication makes use of data products from the Two Micron All
Sky Survey, which is a joint project of the University of
Massachusetts and the Infrared Processing and Analysis
Center/California Institute of Technology, funded by the National
Aeronautics and Space Administration and the National Science
Foundation.  This research was supported by NASA through grant GO1-2119 and 
NSF CAREER award AST-0237291.

\begin{deluxetable}{lllllllllllc}
\rotate
\tabletypesize{\footnotesize}
\tablewidth{0pt}
\tablecaption{Spectral Model Parameters\label{tab:modelpars}}
\tablehead{
\colhead{Region}
&&\colhead{$N$\tablenotemark{a}$_H$}
&\colhead{$kT$\tablenotemark{b}}
&\colhead{$A$\tablenotemark{c}$_{th}$}
&\colhead{$\Gamma$}
&\colhead{$A$\tablenotemark{d}$_{pow}$}
&\colhead{$E$\tablenotemark{e}$_{line}$}
&\colhead{$EW$\tablenotemark{f}$_{line}$}
&\colhead{$f$\tablenotemark{g}$_{0.5-2}$}
&\colhead{$f$\tablenotemark{h}$_{2-10}$}
&\colhead{$\chi^2/$dof}
}
\startdata
Nucleus && 
\nodata
&\nodata&\nodata&\nodata&\nodata&\nodata&\nodata
& $1.7\pm0.3$ 
& $2.1^{+0.5}_{-0.4}$  
& 475/509\tablenotemark{i} \\ 
&AGN&
$1.6\pm0.8$ 
&\nodata&\nodata
& $0.0$f & $0.16^{+0.07}_{-0.06}$  
& $6.40\pm0.05$ & $1.8^{+0.7}_{-0.8}$ 
& $0.04\pm 0.02$ & $1.6^{+0.3}_{-0.4}$ &\nodata\\
\multicolumn{2}{r}{Starburst}& 
$1.6\pm0.8$ 
& $0.08^{+0.01}_{-0:}$ &  $17^{+25}_{-10}$ 
& $1.8$f & $1.4\pm 0.7$  
&\nodata&\nodata
& $1.6^{+0.3}_{-0.2}$ & $0.6\pm0.3$ &\nodata\\

&& \nodata 
& $0.61^{+0.04}_{-0.05}$ & $1.4\pm 0.4$ 
&\nodata&\nodata&\nodata&\nodata&\nodata&\nodata
&\nodata\\
\\
Diffuse
&& 0.2f 
& $0.40\pm 0.05$ & $0.24^{+0.02}_{-0.05}$ 
& $1.8$f & $0.61^{+0.3}_{-0.1}$ 
&\nodata&\nodata
& $0.62\pm 0.06$ 
& $0.21\pm 0.07$
& 40/34
\enddata

\tablenotetext{a}{Column density in units of $10^{21}{\rm\,cm^{-2}}$.}
\tablenotetext{b}{Temperature of thermal plasma in keV.}
\tablenotetext{c}{Normalization of thermal component in units of $10^{-4} K$, 
where $K=(10^{-14}/(4\pi D^2))\int n_e n_H dV, D$ is the distance to the source (cm), 
$n_e$ is the electron density (${\rm cm^{-3}}$), and $n_H$ is the hydrogen density 
(${\rm cm^{-3}}$).}
\tablenotetext{d}{Normalization of power law in units of
$10^{-5} {\rm\,photons\, keV^{-1}\,cm^{-2}\,s^{-1}}$ at 1 keV.}
\tablenotetext{e}{Energy of line center in keV.}
\tablenotetext{f}{Equivalent width of line in keV.} 
\tablenotetext{g}{0.5--2.0 keV model flux in units of
$10^{-13}{\rm\,erg\,cm^{-2}\,s^{-1}}$.}
\tablenotetext{h}{2.0--10.0 keV model flux in units of
$10^{-13}{\rm\,erg\,cm^{-2}\,s^{-1}}$.}
\tablenotetext{i}{C-statistic.}
\tablecomments{
Errors are 90\% confidence limits for one interesting parameter.
Parameters that are constrained by hard limits are marked with a colon.
Fixed parameters are marked with f.
}

\end{deluxetable}

\begin{deluxetable}{lrrrrrrr}
\tabletypesize{\footnotesize}
\tablewidth{0pt}
\tablecaption{Additional Sources\label{tab:ptsrc}}
\tablehead{
\colhead{Source}&
\colhead{R.A.}&\colhead{Decl.}
&\colhead{Counts}
&\colhead{Counts}
&\colhead{Counts}
&\colhead{$f_{0.3-8}$}
&\colhead{$L_{0.3-8}$}\\
&\colhead{(J2000)}&\colhead{(J2000)}
&\colhead{(0.3-8 keV)}
&\colhead{(0.3-2 keV)}
&\colhead{(2-8 keV)}
&\colhead{($10^{-15} {\rm \,erg\,cm^{-2}\,s^{-1}}$)}
&\colhead{($10^{38}{\rm \,erg\,s^{-1}}$)}
}
\startdata
1&  21 48 18.8 &  -34 56 57 &   12 & 12&   0 &  2.7 & 16   \\    
2&  21 48 19.4 &  -34 56 56 &   30 & 30&   0 &  3.2 & 18   \\    
3&  21 48 19.6 &  -34 56 58 &   15 & 13&   2 &  3.0 & 17   \\    
4&  21 48 19.7 &  -34 57 07 &   42 & 31&  11 &  6.0 & 34   \\    
5&  21 48 19.8 &  -34 57 17 &   12 & 10&   2 &  1.4 &  8.1 \\    
6&  21 48 20.6 &  -34 57 00 &    8 &  8&   0 &  1.1 &  6.5 \\    
7&  21 48 21.0 &  -34 57 11 &    8 &  8&   0 &  1.5 &  8.7 \\    
\enddata

\tablecomments{Units of right ascension are hours, minutes, and
seconds, and units of declination are degrees, arcminutes, 
and arcseconds.}
\end{deluxetable}


\begin{deluxetable}{lrr}
\tabletypesize{\footnotesize}
\tablewidth{0pt}
\tablecaption{Observed Luminosities\label{tab:lum}}
\tablehead{
\colhead{Region}
&\multicolumn{1}{c}{$L_{0.5-2}$}
&\multicolumn{1}{c}{$L_{2-10}$}\\
&\multicolumn{1}{c}{($10^{40}{\rm\,erg\,s^{-1}}$)}
&\multicolumn{1}{c}{($10^{40} {\rm \,erg\,s^{-1}}$)}
}
\startdata
NGC 7130 (total)   & 15\phantom{.0} & 15\phantom{.0} \\
AGN                & 0.2 & 8.9 \\
Central starburst  & 11\phantom{.0}  & 4.0 \\
Extended starburst & 3.5  & 1.2 \\

\enddata
\end{deluxetable}

\end{document}